\documentclass[10pt,aps,prd,showpacs,twocolumn,groupedaddress]{revtex4}
\usepackage[english]{babel}
\usepackage{amsfonts}
\usepackage{amssymb}
\usepackage{amsmath}
\usepackage{mathrsfs}
\usepackage{hyperref}
\usepackage{graphicx}
\usepackage{upgreek}
\usepackage[mathcal]{euscript}

\date{\today}

\newcommand{\pp}[1]{\left( #1 \right)}
\newcommand{\cc}[1]{\left[ #1 \right]}
\newcommand{\ch}[1]{\left\{ #1 \right\}}
\newcommand{\tr}[1]{\textrm{#1}}
\newcommand{\del}{\partial}

\begin{document}

\title{Quantization of unstable linear scalar fields in static spacetimes}

\author{William C.\ C.\ Lima}
\email{wccl@ift.unesp.br}
\affiliation{Instituto de F\'\i sica Te\'orica, Universidade Estadual Paulista, Rua Dr.\ Bento Teobaldo Ferraz, 271,
Bloco II, CEP 01140-070, S\~ao Paulo, S\~ao Paulo, Brazil}


\begin{abstract}
We discuss the quantization of an unstable field through the construction of a ``one-particle Hilbert space.'' The
system considered here is a neutral scalar field evolving over a globally hyperbolic static spacetime and subject to a
stationary external scalar potential. In order to prove our results we assume spacetimes without horizons and that the
theory possess a ``mass gap.'' Our strategy consists in building a complex structure, which arises from a suitable
positive bilinear form defined over the space of classical solutions of the field equation. Once the space of states of
the quantum field has been set, it is possible to study the effect of the time translation symmetry on it. From the time
translation operator we obtain an expression for the Hamiltonian operator associated with the unstable sector of the
field. This last result coincides with findings from long ago showing that the unstable degrees of freedom of the field
behave as non-relativistic particles in a parabolic potential barrier.   
\end{abstract}

\maketitle

%%%%%%%%%%%%%%%%%%%%%%%%%%%%%%%%%%%%%%%%%%%%%%%%%%%%%%%%%%%%%%%%%%%%%
\section{Introduction}
\label{introduction}
%%%%%%%%%%%%%%%%%%%%%%%%%%%%%%%%%%%%%%%%%%%%%%%%%%%%%%%%%%%%%%%%%%%%%

One of the most interesting features of quantum fields is the instabilities that may arise when they interact with some
external classical field. Here we are particularly concerned with the tachyonic-like instability analyzed in Ref.\
\cite{lima_and_vanzella_2010}. There it was argued that it is possible for a neutral, non-minimally coupled
scalar field to become unstable when subject to gravitational fields generated by realistic classical-matter
distributions. The presence of the instability leads to an exponential amplification of the vacuum fluctuations of the
field and, consequently, of the expected value of its energy-momentum tensor. A concrete realization of this
claim was given in Ref.\ \cite{lima_matsas_vanzella_2010}, where the appearance of the instability was studied in the
spacetime of a compact object. The scenario in which the spacetime is generated by a spheroidal shell was considered in
Ref.\ \cite{lima_mendes_matsas_vanzella_2013}. This last example is a good toy model to analyze the relation
between the instability and deviations from the spherical symmetry. Obviously, this instability cannot persist
indefinitely: the growth of the observables related to the field must be detained by its backreaction on the background
spacetime. The final state of the system is presently under debate \cite{pani_et_al_2011}. However, whatever it might
be, it is reasonable to expect that --- possibly important --- particle creation processes will occur while the system
reaches its new stable configuration \cite{landulfo_lima_matsas_vanzella_2012}. 

In Minkowski spacetime this kind of instability was first studied by Schiff, Snyder, and Weinberg (SSW) in a paper
published in 1940 \cite{schiff_snyder_weinberg_1940}. In that paper the authors considered a charged scalar field
subject to an electrostatic potential well. For sufficiently deep wells, they discovered that the field equation allows
modes with ``complex frequencies.'' Their main concern, however, was about the particle interpretation of these modes.
This interpretation becomes difficult since the Hamiltonian is no longer related to the number operator. (For an
extensive study on SSW and other instabilities in this model, see Ref.\ \cite{fulling_1976}.) The discussion on the
quantum field-theoretic treatment of the SSW instability was deepened 30 years later by Schr\"oer and Swieca by
considering the simpler problem of a charged scalar field with an external scalar interaction
\cite{schroer_and_swieca_1970}. In their approach they concluded that the field algebra can be represented either in a
space with an indefinite metric or in a Hilbert space without a ground state. Moreover, their analysis of the field
instability revealed that each unstable degree of freedom behaves as a non-relativistic quantum particle in a parabolic
potential barrier.

In this paper we revisit the quantization of fields presenting tachyonic-like instabilities over globally hyperbolic
static backgrounds on the light of the formalism developed for free quantum fields in curved spacetimes
\cite{fulling_book, wald_red_book}. The strategy we shall employ consists in defining a particular complex structure
over the space of solutions of the field equation in order to select a ``one-particle Hilbert space.'' The
representation for the algebra of the unstable field can then be constructed through a Fock space over the chosen
``one-particle Hilbert space.'' To put forward this strategy we shall restrict ourselves to background spacetimes that
do not possess horizons and will assume that the theory possess a ``mass gap.''

In the context of quantum field theory in curved spacetimes (QFTCS), this approach was first introduced by Ashtekar and
Magnon in Ref.\ \cite{ashtekar_and_magnon_1975} and further developed by Kay in Ref.\ \cite{kay_1978} for stable fields
in stationary spacetimes without horizons. Curiously enough, the quantization of unstable fields presenting
tachyonic-like instabilities along these lines seems to be lacking in the literature; here we cover this gap. Our
starting point will be a result due to Chmielowski \cite{chmielowski_1994}, which states that to every bilinear form
over the space of classical solutions defining a quasifree state on the field algebra there is associated a different
bilinear form defining a pure quasifree state.

It is important to emphasize that the canonical quantization procedure does not rely on any specific choice for the
representation of the field algebra. Nevertheless, a particular representation may display more clearly some features of
the system. In the case we are interested in, we desire a representation for the field algebra that does not
``mix'' the stable degrees of freedom of the field with the unstable ones. Since the backreaction of the quantum field
is expected to be dominated by the instability, the separated analysis of the unstable sector and its time evolution
might provide us some clues about this process. Here, however, we shall not be concerned about backreaction effects;
this issue is far beyond the scope of this study. 

This paper is organized as follows. In Sec.\ \ref{sec:standard_quantization_general_spacetimes} we follow Refs.\
\cite{chmielowski_1994,wald_red_book} and present the general theoretical framework to define a representation for the
field algebra based on a ``one-particle Hilbert space.'' In Sec.\ \ref{sec:unstable_one_particle_hilbert_space} we
restrict our attention to unstable fields over static spacetimes without horizons and fix our bilinear form. Once the
representation has been set, it is possible to implement the time-translation symmetry of the background spacetime as a
one-parameter unitary group over our Fock space. To do so, we first deal with the time translation for the
``one-particle states'' in Sec.\ \ref{sec:time_translation_in_H}, trying to keep the analysis for the stable sector as
close as possible to previous ones. Then, in Sec.\ \ref{sec:time_translation_in_F}, we move to the definition of the
time translation on the Fock space. As will be shown, this problem can be treated using the formalism presented in Ref.\
\cite{wald_1979} to define the $S$-matrix for free fields in curved spacetimes. Next, we analyze in Sec.\
\ref{sec:hamiltonian} the Hamiltonian operator associated with the quantum system and ``rediscover'' the connection
between unstable fields and non-relativistic quantum particles in parabolic potential barriers. We close the discussion
and make our final remarks in Sec.\ \ref{sec:conclusion}. Throughout the text we shall assume natural units
($G=\hbar=c=1$) and the signature $(-+++)$ for the spacetime metric.

%%%%%%%%%%%%%%%%%%%%%%%%%%%%%%%%%%%%%%%%%%%%%%%%%%%%%%%%%%%%%%%%%%%%%
\section{Free field quantization in globally hyperbolic spacetimes}
\label{sec:standard_quantization_general_spacetimes}
%%%%%%%%%%%%%%%%%%%%%%%%%%%%%%%%%%%%%%%%%%%%%%%%%%%%%%%%%%%%%%%%%%%%%

\subsection{Classical field}
%%%%%%%%%%%%%%%%%%%%%%%%%%%%%%%%%%%%%%%%%%%%%%%%%%

For a globally hyperbolic spacetime $(\mathcal{M},g_{ab})$ it is always possible to choose a foliation
$\{\Sigma_t\}_{t\in\mathbb{R}}$ where each hypersurface $\Sigma_t$ is a smooth space-like Cauchy surface
\cite{bernal_and_sanchez_2003}. Denote by $n^a$ the normalized time-like vector field orthogonal to every hypersurface
in the foliation. The induced metric defined over each spatial section is given by 
\begin{equation*}
 h_{ab}\equiv n_an_b+g_{ab}.
\end{equation*}
We can introduce a coordinate system in our spacetime by first setting the time-like vector field
$t^a$ on $\mathcal{M}$ in such way that $t^a\nabla_a t=1$. Then, we cover each hypersurface $\Sigma_t$ with coordinates
$x^i$
($i=1,2,3$) satisfying $t^a\nabla_ax^i=0$, so $t^a$ is our time coordinate vector.

Here we shall consider a neutral scalar field over $(\mathcal{M},g_{ab})$ with an external scalar interaction
defined by the following action:
\begin{equation}\label{field_action}
S\equiv-\frac{1}{2}\int_{\mathcal{M}}{\sqrt{-g}d^4x\pp{\nabla_a\phi\nabla^a\phi+V\phi^2}},
\end{equation}
with $V$ a real function over $\mathcal{M}$ (not explicitly depending on $\phi$), which could encompass terms like mass
and non-minimal coupling with the scalar curvature. The principle of minimal action applied to Eq.\ (\ref{field_action})
leads to the following field equation:
\begin{equation}\label{kg_eq}
-\nabla_a\nabla^a\phi+V\phi=0,
\end{equation} 
the Klein-Gordon equation with an external scalar potential. The hypothesis of global hyperbolicity implies that the
specification of a ``sufficiently nice'' pair of functions $(f,p)$ in a Cauchy surface $\Sigma_0$ such that
$f=\phi|_{\Sigma_0}$ and $p=n^a\nabla_a\phi|_{\Sigma_0}$ --- the initial conditions --- determines univocally a solution
of Eq.\ (\ref{kg_eq}) over all the spacetime (see, e.g., Theorem 5.3.1 of Ref.\ \cite{friedlander_book}). 

For general spacetimes, the convenient choice for the phase space $\mathcal{P}$ of the field is
\begin{equation*}
\mathcal{P}\equiv\ch{(f,p)|f, p:\Sigma_0\to\mathbb{R}; f, p\in C^\infty_0(\Sigma_0)},
\end{equation*}
where $C^\infty_0(\Sigma_0)$ stands for the set of smooth real functions with compact support in $\Sigma_0$. Let us
assume, then, that the background spacetime and the scalar potential $V$ are smooth. According to Leray's theorem
\cite{friedlander_book,leray_notes}, in this case $\phi$ and $n^a\nabla_a\phi$ are also smooth functions and of compact
support when restricted to any hypersurface $\Sigma_t$. The real vector space $\mathcal{S}$ is then defined as the set
of all solutions of Eq.\ (\ref{kg_eq}) with initial data in $\mathcal{P}$. The fact that the Cauchy problem for Eq.\
(\ref{kg_eq}) has a unique solution for a given initial data implies that the vector spaces $\mathcal{P}$ and
$\mathcal{S}$ are isomorphic. Hence, we can refer to these spaces interchangeably. 

Over the phase space $\mathcal{P}$ it is possible to define a bilinear form
$\Omega:\mathcal{P}\times\mathcal{P}\rightarrow\mathbb{R}$ by  
\begin{equation}\label{symplectic_form}
\Omega(\psi_1,\psi_2)\equiv\int_{\Sigma_0}{d\Sigma\pp{\phi_1n^a\nabla_a\phi_2-\phi_2n^a\nabla_a\phi_1}},
\end{equation}
where $\psi$ stands for the pair $(\phi,n^a\nabla_a\phi)$ and $d\Sigma$ denotes the volume element of $\Sigma_t$, even
though calculated over a given Cauchy surface, the right-hand side of Eq.\ (\ref{symplectic_form}) is conserved by the
dynamics defined by Eq.\ (\ref{kg_eq}). This fact, together with the existing isomorphism between $\mathcal{P}$ and
$\mathcal{S}$, induces a bilinear form over $\mathcal{S}$, which is independent of the Cauchy surface $\Sigma_0$ as
well. We also denote this bilinear form by $\Omega$. Since it is antisymmetric and non-degenerated for solutions of Eq.\
(\ref{kg_eq}), $\Omega$ is said to provide a symplectic structure on $\mathcal{S}$. As it will become clear in the next
section, the only structure we need for the quantization of a neutral scalar field is the real symplectic vector space
$(\mathcal{S},\Omega)$.

\subsection{Quantum field}
%%%%%%%%%%%%%%%%%%%%%%%%%%%%%%%%%%%%%%%%%%%%%%%%%%

To quantize a linear scalar field one seeks a complex, separable Hilbert space $\mathcal{F}$ and a map $W$ which takes
the elements of $\mathcal{S}$ into unitary operators acting on $\mathcal{F}$ satisfying
\begin{equation}\label{weyl_1}
W(\psi_1)W(\psi_2)=e^{\frac{i}{2}\Omega(\psi_1,\psi_2)}W(\psi_1+\psi_2)
\end{equation}
as well as 
\begin{equation}\label{weyl_2}
W^\dagger(\psi)=W(-\psi),
\end{equation}
the Weyl relations. From the technical point of view, it is safer to deal with these unitary operators than with the
field operators satisfying the canonical commutation relations since the formers are bounded. 

The standard way to approach this question is to start with a pair $(K,\mathcal{H})$, where $\mathcal{H}$ is a complex,
separable Hilbert space and $K:\mathcal{S}\to\mathcal{H}$ is a real, linear map such that its range $\textrm{Ran}(K)$ is
dense in $\mathcal{H}$. The Hilbert space $\mathcal{F}$ is then defined as a Fock space built out of $\mathcal{H}$.
Hence, inasmuch as we are concerned with bosonic quantum fields, we define $\mathcal{F}$ as a symmetric Fock space,
i.e.,
\begin{equation}\label{fock_space}
\mathcal{F}=\mathfrak{F}_\textrm{s}(\mathcal{H})
\equiv\mathbb{C}\oplus\mathcal{H}\oplus(\mathcal{H}\otimes\mathcal{H})_\textrm{s}\oplus\dots,
\end{equation}
where $(\bigotimes^n\mathcal{H})_\textrm{s}$ denotes the symmetric subspace of $\bigotimes^n\mathcal{H}$. The
inner product of $\mathcal{F}$ is defined through 
\begin{equation*}
\langle\Psi,\Phi\rangle_\mathcal{F}\equiv\overline{c}d+\sum_{n=1}^\infty\langle\psi^{(n)},\varphi^{(n)}
\rangle_{\otimes^n\mathcal{H}},
\end{equation*}
with $c,d\in \mathbb{C}$, $\bar{c}$ the complex conjugate of $c$,
$\varphi^{(n)},\psi^{(n)}\in(\bigotimes^n\mathcal{H})_\textrm{s}$, and $\langle\; \cdot\; ,\;
\cdot\; \rangle_{\otimes^n\mathcal{H}}$ denoting the inner product of $\bigotimes^n\mathcal{H}$. 

Consider, then, $\mathcal{F}_0\subset\mathcal{F}$, the set of all $n$-particle states, i.e., the set of all vectors of
$\mathcal{F}$ with the form
\begin{equation*}
\Psi=\pp{0,\dots,0,\psi^{(n)},0,\dots},
\end{equation*}
where $\psi^{(n)}\in(\bigotimes^n\mathcal{H})_\textrm{s}$, for $n\in\mathbb{N}$. Then, the linear span of
$\mathcal{F}_0$, $\textrm{span}(\mathcal{F}_0)$, is a dense subset of $\mathcal{F}$. Next, for
$\Psi\in\textrm{span}(\mathcal{F}_0)$,
$\chi\in\mathcal{H}$, and $\overline{\chi}\in\overline{\mathcal{H}}$ --- the complex conjugate space to $\mathcal{H}$
---, one defines the operator
\begin{equation}\label{annihilation_operator}
a(\overline{\chi})\Psi\equiv\pp{\overline{\chi}\cdot\psi^{(1)},\sqrt{2}\overline{\chi}\cdot\psi^{(2)},
\sqrt{3}\overline{\chi}\cdot\psi^{(3)},\dots }
\end{equation}
the annihilation operator, and
\begin{equation}\label{creation_operator}
a^*(\chi)\Psi\equiv\pp{0,c\chi,\sqrt{2}\chi\otimes_\textrm{s}\psi^{(1)},
\sqrt{3}\chi\otimes_\textrm{s}\psi^{(2)},\dots}
\end{equation}
the creation operator. The dot that appears in the slots in the right-hand side of Eq.\ (\ref{annihilation_operator})
indicates the contraction, according to the inner product of $\mathcal{H}$, between the vectors $\chi$ and $\psi^{(n)}$,
while $\otimes_\textrm{s}$ in Eq.\ (\ref{creation_operator}) indicates the symmetrized tensor product. These operators
are continuous in the sense that if $\chi_n\rightarrow\chi$ in $\mathcal{H}$ then $a(\overline{\chi}_n)\Psi \rightarrow
a(\overline{\chi})\Psi$ and $a^*(\chi_n)\Psi\rightarrow a^*(\chi)\Psi$ in $\mathcal{F}$. For future reference, we
denote by $\Psi_0$ the normalized element in $\mathcal{F}_0$ satisfying
\begin{equation}\label{vacuum_state}
a(\overline{\chi})\Psi_0=0
\end{equation}
for all $\chi\in\mathcal{H}$, the vacuum state. From Eqs.\ (\ref{annihilation_operator}) and (\ref{creation_operator})
it is easy to verify that for any vector in $\textrm{span}(\mathcal{F}_0)$ and any $\chi_1,\chi_2 \in \mathcal{H}$
\begin{equation}\label{annihilation_commutator}
[a(\overline{\chi}_1),a(\overline{\chi}_2)]=0,
\end{equation}
\begin{equation}\label{creation_commutator}
[a^*(\chi_1),a^*(\chi_2)]=0,
\end{equation}
and
\begin{equation}\label{annihilation_creation_commutator}
[a(\overline{\chi}_1),a^*(\chi_2)]=\langle \chi_1,\chi_2\rangle_\mathcal{H} \mathbb{I},
\end{equation}
where $\mathbb{I}$ stands for the identity operator.

With the aid of the creation and annihilation operators one defines the field operator for any $\psi\in\mathcal{S}$ as
\begin{equation}\label{field_operator}
F_0(\psi)\equiv ia(\overline{K\psi})-ia^*(K\psi).
\end{equation}
The field operator $F_0(\psi)$ is essentially self-adjoint in $\textrm{span}(\mathcal{F}_0)$, so it has just one
self-adjoint extension, namely, its closure $F(\psi)$. The representation for the relations (\ref{weyl_1}) and
(\ref{weyl_2}) is then set when we take $F(\psi)$ as the generator of $W(\psi)$:
\begin{equation*}
W(\psi)\equiv e^{iF(\psi)}.
\end{equation*}
Sometimes in the literature the quantization scheme presented here is referred to as the ``Segal quantization over
$\mathcal{H}$'' \cite{segal_1962,weinless_1969}. For the proof of the continuity of the creation and annihilation
operators, the essentially self-adjointness of $F_0(\psi)$, and other properties of the Segal quantization see, e.g.,
Theorem X.41 of Ref.\ \cite{reed_and_simon_v2}.

Note that the vacuum state defines the following functional:
\begin{equation*}
\langle\Psi_0,W(\psi)\Psi_0\rangle_\mathcal{F}=e^{-\frac{1}{2}\langle K\psi,K\psi\rangle_\mathcal{H}}. 
\end{equation*}
From the algebraic point of view, $\Psi_0$ induces a state over the $\textrm{C}^*$-algebra defined by the operators
$W(\psi)$. (For a comprehensive discussion on the algebraic formalism for quantum fields, see, e.g., Ref.\
\cite{haag_book}.)

\subsection{Definition of the ``one-particle structure''}
\label{sec:one_particle_structure}
%%%%%%%%%%%%%%%%%%%%%%%%%%%%%%%%%%%%%%%%%%%%%%%%%% 

The central question here, then, is how to define the pair $(K,\mathcal{H})$. A possible route starts by considering a
bilinear, positive, symmetric form $\mu$ on $\mathcal{S}$ satisfying
\begin{equation}\label{bounded_symplectic_structure_mu}
[\Omega(\psi_1,\psi_2)]^2\le4\mu(\psi_1,\psi_1)\mu(\psi_2,\psi_2)
\end{equation}
for any $\psi_1,\psi_2\in\mathcal{S}$. Denote by $\mathcal{A}$ the real Hilbert space obtained by the completion of
$\mathcal{S}$ in the norm defined by $\mu$. Then, on $\mathcal{A}$, $\Omega$ is a bounded, densely defined bilinear form
--- the boundedness of the symplectic form is actually the reason of Eq. (\ref{bounded_symplectic_structure_mu}). Hence,
by the bounded linear transformation theorem (see, e.g., Theorem I.7 of Ref.\ \cite{reed_and_simon_v1}), there is a
unique bounded extension of $\Omega$ to $\mathcal{A}$, which we denote by $\Omega^\prime$. The boundedness of
$\Omega^\prime$ implies that on $\mathcal{A}$ there exists a bounded operator $A$ such that
\begin{equation}\label{A_operator_definition}
\mu(\psi_1,A\psi_2)\equiv\frac{1}{2}\Omega^\prime(\psi_1,\psi_2).
\end{equation}
It follows from Eq.\ (\ref{A_operator_definition}) and the antisymmetry property of $\Omega$ that the operator $A$ is
also antisymmetric. 

The polar decomposition theorem for bounded operators (see, e.g., Theorem VI.10 of Ref.\
\cite{reed_and_simon_v1}) allows us to write $A=U|A|$, where $|A|$ is the unique positive operator such that
$|A|^2=A^\dagger A=-A^2$ and $U$ is a partial isometry. The non-degeneracy of $\Omega$ implies for the kernel of $A$
that $\mathcal{S}\cap\textrm{Ker}(A)=\{0\}$. Consequently, the following expression defines a bilinear, positive,
symmetric form over the space of solutions $\mathcal{S}$:
\begin{equation}\label{new_bilinear_form}
\tilde{\mu}(\psi_1,\psi_2)\equiv\mu(\psi_1,|A|\psi_2).
\end{equation}
Then, it is possible to show (see Proposition 1 of Ref.\ \cite{chmielowski_1994}) that
\begin{equation}\label{bounded_symplectic_form}
\tilde{\mu}(\psi_1,\psi_1)\tilde{\mu}(\psi_2,\psi_2)\ge\frac{1}{4}[\Omega(\psi_1,\psi_2)]^2
\end{equation}
and that
\begin{equation}\label{lub_mu_tilde}
\tilde{\mu}(\psi_1,\psi_1)=\frac{1}{4}
\operatorname*{l.u.b.}_{\psi_2\neq0}\frac{[\Omega(\psi_1,\psi_2)]^2}{\tilde{\mu}(\psi_2,\psi_2)},
\end{equation}
where $\operatorname*{l.u.b.}$ stands for ``least upper bound.'' In the algebraic approach for quantum fields, a
bilinear form defined over $\mathcal{S}$ satisfying Eq.\ (\ref{bounded_symplectic_structure_mu}) defines a
quasifree state, while Eq.\ (\ref{lub_mu_tilde}) gives rise to a pure quasifree state \cite{kay_and_wald_1991}. 

Next, define $\tilde{\mathcal{A}}$ as the real Hilbert space resulting from the completion of $\mathcal{S}$ in the norm
defined by the following inner product:
\begin{equation}\label{A_tilde_inner_product}
\langle\psi_1,\psi_2\rangle_{\tilde{\mathcal{A}}}\equiv2\tilde{\mu}(\psi_1,\psi_2).
\end{equation}
From Eqs.\ (\ref{A_operator_definition}) and (\ref{new_bilinear_form}) it is easy to see that for
$\psi_1,\psi_2\in\mathcal{S}$, the partial isometry $U$ satisfies
\begin{equation}\label{U_isometry}
\tilde{\mu}(U\psi_1,U\psi_1)=\tilde{\mu}(\psi_1,\psi_1)
\end{equation}
and
\begin{equation}\label{U_antisymmetry}
\tilde{\mu}(\psi_1,U\psi_2)=-\tilde{\mu}(U\psi_1,\psi_2).
\end{equation} 
Equation (\ref{U_isometry}) states that $U$ is an isometry according to this norm, so we can extend its action to
$\tilde{\mathcal{A}}$. Denote this extension by $J$. Due to Eq.\ (\ref{U_antisymmetry}), the operator $J$
satisfies
\begin{equation}\label{J_antisymmetry}
J^\dagger=-J.
\end{equation}
The fact that $J$ is an isometry, together with its antisymmetry, implies that 
\begin{equation}\label{J_square}
J^2=-\mathbb{I}.
\end{equation}
A linear operator $J$ satisfying Eqs.\ (\ref{J_antisymmetry}) and (\ref{J_square}) is said to endow the real Hilbert
space $\tilde{\mathcal{A}}$ with a complex structure \cite{footnote_1}. The symplectic form $\Omega$ can also be
extended from $\mathcal{S}$ to $\tilde{\mathcal{A}}$ thanks to Eq.\ (\ref{bounded_symplectic_form}). This extension will
be denoted by $\Omega^{\prime\prime}$. Due to Eqs.\ (\ref{A_operator_definition}) and (\ref{new_bilinear_form}), it is
related to $J$ by
\begin{equation}\label{J_sigma}
\langle\psi_1,J\psi_2\rangle_{\tilde{\mathcal{A}}}=\Omega^{\prime\prime}(\psi_1,\psi_2).
\end{equation}

The next step consists in complexifying $\tilde{\mathcal{A}}$, resulting in the space $\tilde{\mathcal{A}}_\mathbb{C}$,
and extending to it by complex linearity the bilinear forms $\tilde{\mu}$, $\Omega^{\prime\prime}$, and the operator
$J$. Naturally, the space $\tilde{\mathcal{A}}_\mathbb{C}$ is a complex Hilbert space in the norm associated with the
following inner product:
\begin{equation}\label{inner_product}
\langle\psi_1,\psi_2\rangle_{\tilde{\mathcal{A}}_\mathbb{C}}\equiv2\tilde{\mu}(\overline{\psi}_1,\psi_2).
\end{equation}

Now we are in a position to define the Hilbert space $\mathcal{H}$ as a closed subspace of
$\tilde{\mathcal{A}}_\mathbb{C}$. To do so, note that $iJ$ is self-adjoint and, due to Eq.\ (\ref{J_square}), has
eigenvalues $\pm 1$. Therefore, from the spectral theorem, $\tilde{\mathcal{A}}_\mathbb{C}$ decomposes into two
orthogonal eigensubspaces of $iJ$ which are complex conjugates of each other. Denote by $P_\pm$ the orthogonal
projections associated with the eigenvalues $\pm1$. Then,
\begin{equation}\label{J_spectral_form}
J=-i(P_+-P_-)
\end{equation}
and the ``one-particle Hilbert space'' $\mathcal{H}$ is defined as $\textrm{Ran}(P_+)$. For any two vectors
$\psi_1,\psi_2\in\tilde{\mathcal{A}}_\mathbb{C}$ it is easy to show that 
\begin{eqnarray}\label{kg_inner_product}
\langle P_+\psi_1,P_+\psi_2\rangle_\mathcal{H}&\equiv&\langle
P_+\psi_1,P_+\psi_2\rangle_{\tilde{\mathcal{A}}_\mathbb{C}}\nonumber\\
&=&i\Omega^{\prime\prime}(\overline{P_+\psi}_1,P_+\psi_2),
\end{eqnarray}
usually called ``Klein-Gordon inner product.'' The definition of $\mathcal{H}$ implies that
$\overline{\mathcal{H}}=\textrm{Ran}(P_-)$, i.e., $\overline{\mathcal{H}}$ corresponds to the orthogonal
complement of $\mathcal{H}$ in $\tilde{\mathcal{A}}_\mathbb{C}$.

The restriction of the projection $P_+$ to $\mathcal{S}$ gives us the real, linear map $K$. As proved
in Ref.\ \cite{kay_and_wald_1991}, Proposition 3.1, the pair $(K,\mathcal{H})$ is uniquely defined, up to a unitary
equivalence. From Eq.\ (\ref{kg_inner_product}) it follows that $K$ satisfies
\begin{eqnarray*}
\langle K\psi_1,K\psi_2\rangle_{\mathcal{H}}=\tilde{\mu}(\psi_1,\psi_2)-\frac{i}{2}\Omega(\psi_1,\psi_2).
\end{eqnarray*}
This relation shows that all the freedom we have in choosing $\mathcal{H}$ is encoded in the bilinear form
$\tilde{\mu}$, which depends on $\mu$ through Eq.\ (\ref{new_bilinear_form}). The freedom in picking a ``one-particle
Hilbert space'' is an important issue in quantum field theory, since different choices can lead, in general, to unitary
inequivalent representations for the Weyl relations.

%%%%%%%%%%%%%%%%%%%%%%%%%%%%%%%%%%%%%%%%%%%%%%%%%%%%%%%%%%%%%%%%%%%%%
\section{Unstable free fields in static spacetimes without horizons}
\label{sec:instability_quantization}
%%%%%%%%%%%%%%%%%%%%%%%%%%%%%%%%%%%%%%%%%%%%%%%%%%%%%%%%%%%%%%%%%%%%%

From now on we shall restrict ourselves to the special case of a static spacetime. Hence, over $(\mathcal{M},g_{ab})$
exists a one-parameter group of isometries $\alpha_t$ generated by a time-like Killing vector field $\varkappa^a$ ---
which we identify with our time coordinate vector $t^a$ --- and a foliation orthogonal to $\varkappa^a$. Thus, we can
choose
\begin{equation*}
n^a=\frac{\varkappa^a}{\|\varkappa\|},
\end{equation*}
where $\|\varkappa\|\equiv\sqrt{-\varkappa^a\varkappa_a}$. Additionally, we will assume that for some Cauchy surface in
the foliation it is possible to find constants $\epsilon_1,\epsilon_2>0$ such that
\begin{equation}\label{bounded_killing_vector_norm}
\epsilon_1<\|\varkappa\|<\epsilon_2.
\end{equation}
Therefore, here we have in mind static spacetimes without horizons. 

When the background spacetime is static, it is possible to cast Eq.\ (\ref{kg_eq}) as
\begin{equation}\label{kg_eq_static}
-\del^2_t\phi=\|\varkappa\|\Delta_0\phi,
\end{equation}
where the differential operator $\Delta_0$ is defined as
\begin{equation}\label{Delta_operator}
\Delta_0 f\equiv-D_a(\|\varkappa\|D^af)+\|\varkappa\|Vf,
\end{equation}
with $D_a$ the covariant derivative operator associated with the spatial metric $h_{ab}$. In order to guarantee that
Eq.\ (\ref{kg_eq_static}) shares the time symmetry of the background spacetime, we will only consider scalar potentials
with null Lie derivative with respect to $\varkappa^a$. If we define the one-parameter family of linear maps
$\tau_t:\mathcal{S}\rightarrow\mathcal{S}$ acting on $\psi=(\phi,n^a\nabla_a\phi)$ as
\begin{equation}\label{time_translation}
\tau_t\psi\equiv\left[
\begin{array}{c}
\phi\circ\alpha_{t}\\
(n^a\nabla_a\phi)\circ\alpha_{t}
\end{array}
\right],
\end{equation}
then the equation
\begin{equation}\label{field_dynamics_matrix_form}
\left.\frac{d}{dt}\tau_t\psi\right|_{t=0}=-\mathbf{h}_0\psi,
\end{equation}
with
\begin{equation}
\mathbf{h}_0\equiv\left[
\begin{array}{cc}
0 & -\|\varkappa\| \\
\Delta_0 & 0
\end{array}
\right],
\end{equation}
just expresses Eq.\ (\ref{kg_eq_static}) in terms of a pair of coupled differential equations.

\subsection{Definition of the ``one-particle Hilbert space'' for an unstable field}
\label{sec:unstable_one_particle_hilbert_space}
%%%%%%%%%%%%%%%%%%%%%%%%%%%%%%%%%%%%%%%%%%%%%%%%%%

We consider the real Hilbert space $L^2(\Sigma_0,d\Sigma)$ --- which we abbreviate from now on just by $L^2$ ---
in the norm associated with the inner product
\begin{equation*}
\langle f_1,f_2\rangle_{L^2}\equiv\int_{\Sigma_0}{d\Sigma f_1f_2}.
\end{equation*}
On this Hilbert space, the multiplication by $\|\varkappa\|$ defines a bounded, positive, self-adjoint operator with
bounded inverse, due to Eq.\ (\ref{bounded_killing_vector_norm}). Moreover, since $V\in C^\infty(\Sigma_0)$ is
locally square-integrable in the measure $d\Sigma$, Eq.\ (\ref{Delta_operator}) corresponds to an unbounded,
symmetric, densely defined operator with domain $\textrm{Dom}(\Delta_0)=C_0^\infty(\Sigma_0)$ and range
$\textrm{Ran}(\Delta_0)=C_0^\infty(\Sigma_0)$. Furthermore, for this class of potentials, it is possible to show that
$\Delta_0$ is also essentially self-adjoint in $\textrm{Dom}(\Delta_0)$ \cite{ikebe_and_kato_1962}. Therefore,
$\Delta_0$ has just one self-adjoint extension in $L^2$, namely, its closure $\Delta$. 

Denote by $\langle\;\cdot\;,\;\cdot\;\rangle_{L^2\oplus L^2}$ the inner product of $L^2\oplus L^2$. We define the real
Hilbert space $\mathcal{N}$ as the completion of $\mathcal{P}$ in the norm induced by the inner product
\begin{equation*}
\langle\psi_1,\psi_2\rangle_{\mathcal{N}}\equiv\langle\psi_1,\mathbf{N} \psi_2\rangle_{L^2\oplus L^2},
\end{equation*}
where
\begin{equation}\label{N_matrix}
\mathbf{N}\equiv\left[
\begin{array}{cc}
\|\varkappa\|^{-1} & 0\\
0 & \|\varkappa\|
\end{array}
\right].
\end{equation}
Note that as sets $\mathcal{N}$ and $L^2\oplus L^2$ coincide, the space $\mathcal{N}$ is isomorphic to $L^2\oplus L^2$
according to the following unitary transformation:
\begin{equation}\label{unitary_transformation}
\psi\in\mathcal{N}\mapsto\mathbf{N}^\frac{1}{2}\psi\in L^2\oplus L^2.
\end{equation}
Moreover, the smoothness of the background spacetime together with Eq.\ (\ref{bounded_killing_vector_norm}) imply
that $\mathbf{N}^\frac{1}{2}$ maps $\mathcal{P}$ into itself.

Consider, then, the following bilinear, symmetric form over $\mathcal{S}$:
\begin{eqnarray}\label{conserved_energy}
\varepsilon(\psi_1,\psi_2)&\equiv&\frac{1}{2}\int_{\Sigma_0}{d\Sigma(\|\varkappa\|p_1p_2+f_1\Delta_0f_2)}\nonumber\\
&=&\langle\mathbf{N}^{\frac{1}{2}}\psi_1,\mathbf{M}_0\mathbf{N}^{\frac{1}{2}}\psi_2\rangle_{L^2\oplus L^2},
\end{eqnarray}
with
\begin{equation}\label{M_0_matrix}
\mathbf{M}_0\equiv\frac{1}{2}\left[
\begin{array}{cc}
\|\varkappa\|^{\frac{1}{2}}\Delta_0\|\varkappa\|^{\frac{1}{2}} & 0\\
0 & 1
\end{array}
\right],
\end{equation}
where we have used the unitary transformation (\ref{unitary_transformation}). When $\psi_1=\psi_2$, the bilinear form
(\ref{conserved_energy}) is equal to the conserved energy associated with the scalar field. From Eq.\
(\ref{conserved_energy}) we see that the energy of the field can become as negative as we want if $\Delta_0$ fails to be
a positive operator. Therefore, the classical scalar field becomes unstable when $\sigma(\Delta)$, the spectrum of
$\Delta$, acquires a negative part.

From this point on we shall focus on situations in which the scalar field is destabilized by the external fields. Hence,
it will be implied that $\sigma(\Delta) \cap \mathbb{R}^-\neq\emptyset$. For technical reasons, we shall assume the
existence of constants $M_1,M_2>0$ such that for all $\lambda\in\sigma(\Delta)$,
\begin{equation}\label{spectral_condition_1}
|\lambda|>M_1
\end{equation}
and
\begin{equation}\label{spectral_condition_2}
\lambda>-M_2.
\end{equation}
Condition (\ref{spectral_condition_1}) guarantees that the inverse of some operators that will appear in what follows
are bounded. (In the case of stable fields, this imposition is related to the infrared behavior of the theory; for a
discussion on this issue for stable fields see, e.g., Ref.\ \cite{fulling_and_ruijsenaars_1987}.) Moreover, this
condition will be important to prove that Eq.\ (\ref{bounded_symplectic_structure_mu}) holds for our choice for $\mu$
--- see Appendix \ref{appendix}. Equation (\ref{spectral_condition_1}) is what is meant by ``mass gap'' in Sec.\
\ref{introduction}. As for condition (\ref{spectral_condition_2}), it is equivalent to the statement that the scalar
potential $V$ is bounded from below, which is a physically reasonable assumption.

The operator $\mathbf{M}_0$ is essentially self-adjoint on $\mathcal{P}$, since $\|\varkappa\|^{\frac{1}{2}}\Delta_0
\|\varkappa\|^{\frac{1}{2}}$ is essentially self-adjoint on $C_0^\infty(\Sigma_0)$. Denote by $\mathbf{M}$ the closure
of $\mathbf{M}_0$. The map $\|\varkappa\|^{-\frac{1}{2}}\oplus\|\varkappa\|^{\frac{1}{2}}:L^2\oplus
L^2\rightarrow L^2\oplus L^2$ applied to the elements of the graph of the operator $\Delta_0$, 
\begin{equation*}
 \Gamma(\Delta_0)\equiv\ch{(f,\Delta_0f),f\in \textrm{Dom}(\Delta_0)},
\end{equation*}
gives
\begin{equation*}
(f,\Delta_0f)\mapsto(\|\varkappa\|^{-\frac{1}{2}}f,\|\varkappa\|^{\frac{1}{2}}\Delta_0
\|\varkappa\|^{\frac{1}{2}}\|\varkappa\|^{-\frac{1}{2}}f),
\end{equation*}
so $\Gamma(\Delta_0)$ is mapped into $\Gamma(\|\varkappa\|^{\frac{1}{2}}\Delta_0 \|\varkappa\|^{\frac{1}{2}})$. The map
$\|\varkappa\|^{-\frac{1}{2}}\oplus\|\varkappa\|^{\frac{1}{2}}$ is a homeomorphism in the norm topology.
Hence, we also have that the graph of $\Delta$ is taken into the graph of the closure of
$\|\varkappa\|^{\frac{1}{2}}\Delta_0\|\varkappa\|^{\frac{1}{2}}$, which must be equal to
$\|\varkappa\|^{\frac{1}{2}}\Delta\|\varkappa\|^{\frac{1}{2}}$. With the definition
\begin{equation*}
\Lambda\equiv \|\varkappa\|^{\frac{1}{2}}\Delta\|\varkappa\|^{\frac{1}{2}},
\end{equation*}
this last result implies that 
\begin{equation}\label{M_matrix}
\mathbf{M}=\frac{1}{2}\left[
\begin{array}{cc}
\Lambda & 0\\
0 & 1
\end{array}
\right].
\end{equation}
Note that if $\Delta$ fails to be a positive operator, so does $\Lambda$.

For a stable field, a suitable choice for $\mu$ is the bilinear form associated with the conserved energy, Eq.\
(\ref{conserved_energy}). In this case, the field quantization is formally equivalent to the prescription of expanding
the field operator in terms of positive and negative-frequency modes of the field equation
\cite{ashtekar_and_magnon_1975,kay_1978,wald_red_book}. The presence of the instability, however, spoils the positivity
of the energy and, consequently, $\varepsilon$ does not define a norm on $\mathcal{S}$. For the unstable field, then, we
consider the following bilinear, positive, symmetric form \cite{wccl_phd_thesis}:
\begin{equation}\label{mu_definition}
\mu(\psi_1,\psi_2)\equiv\langle\mathbf{N}^{\frac{1}{2}}\psi_1,|\mathbf{M}|\mathbf{N}^{\frac{1}{2}}\psi_2\rangle_{
L^2\oplus L^2},
\end{equation}
with
\begin{equation}\label{|M|_matrix}
|\mathbf{M}|=\frac{1}{2}\left[
\begin{array}{cc}
|\Lambda| & 0\\
0 & 1
\end{array}
\right].
\end{equation}
As shown in Appendix \ref{appendix}, the bilinear form (\ref{mu_definition}) satisfies Eq.\
(\ref{bounded_symplectic_structure_mu}), so we can proceed with the formalism discussed in Sec.\
\ref{sec:standard_quantization_general_spacetimes}. What is interesting about the bilinear form defined by Eq.\
(\ref{mu_definition}) is that it coincides with the conserved energy (\ref{conserved_energy}) whenever the latter is
positive. Hence, for this stable sector, the bilinear form (\ref{mu_definition}) is conserved and the quantization that
arises when we follow the steps presented in Sec.\ \ref{sec:standard_quantization_general_spacetimes} does not rely on
the choice of the Cauchy surface. For the unstable sector, however, $\mu$ is not conserved. Thus, in general, our
definition of $\mathcal{H}$ will depend on the Cauchy surface on which we give the initial conditions for the field. As
will be discussed in Sec.\ \ref{sec:time_translation_in_H}, this fact implies that the time translation defined by Eq.\
(\ref{time_translation}) will not induce a unitary operator on $\mathcal{H}$.

Once the bilinear form $\mu$ has been fixed, it is possible to give a representation for the Hilbert space
$\mathcal{A}$, for the bounded operator $A$, and its polar decomposition. One sees from the unitary transformation
(\ref{unitary_transformation}) and Eq.\ (\ref{mu_definition}) that $\mathcal{A}$ is isomorphic to $\mathcal{A}_1\oplus
L^2$, where $\mathcal{A}_1$ is the Hilbert space resulting from the completion of $C_0^\infty(\Sigma_0)$ in the norm
defined by $\frac{1}{2}\langle\;\cdot\;, |\Lambda| \;\cdot\;\rangle_{L^2}$. Then, the action of the operator $A$ can be
written as
\begin{equation}\label{A_representation}
\mu(\psi_1,A\psi_2)=\langle\mathbf{N}^{\frac{1}{2}}\psi_1,|\mathbf{M}|\check{A}
\mathbf{N}^{\frac{1}{2}}\psi_2\rangle_{L^2\oplus L^2},
\end{equation}
with
\begin{equation}\label{A_unitary_transformation}
\check{A}\equiv\mathbf{N}^\frac{1}{2}A\mathbf{N}^{-\frac{1}{2}}.
\end{equation}
Using the inner product of $L^2\oplus L^2$, we can express the symplectic form (\ref{symplectic_form}) as 
\begin{equation}\label{matrix_symplectic_form}
\Omega(\psi_1,\psi_2)=\langle\psi_1,\mathbf{g}\psi_2\rangle_{L^2\oplus L^2},
\end{equation}
with
\begin{equation}\label{g_matrix}
\mathbf{g}\equiv\left[
\begin{array}{cc}
0 & 1\\
-1 & 0
\end{array}
\right].
\end{equation}
Then, comparing Eqs.\ (\ref{A_representation}) and (\ref{matrix_symplectic_form}) as in Eq.\
(\ref{A_operator_definition}), it follows that
\begin{equation}\label{A_matrix}
\check{A}=\frac{1}{2}|\mathbf{M}|^{-1}\mathbf{N}^{-\frac{1}{2}}\mathbf{g}\mathbf{N}^{-\frac{1}{2}}
=\left[
\begin{array}{cc}
0 & |\Lambda|^{-1}\\
-1 & 0
\end{array}
\right].
\end{equation}
Note that $|\Lambda|^{-1}$ exists as a positive, bounded operator, due to property (\ref{spectral_condition_1}) of
$\sigma(\Delta)$ and Eq.\ (\ref{bounded_killing_vector_norm}). For the unitary transformation of the operator $|A|$ one
has
\begin{equation}\label{|A|_matrix}
\check{|A|}\equiv\mathbf{N}^{\frac{1}{2}}|A|\mathbf{N}^{-\frac{1}{2}}=\left[
\begin{array}{cc}
|\Lambda|^{-\frac{1}{2}} & 0\\
0 & |\Lambda|^{-\frac{1}{2}}
\end{array}
\right].
\end{equation}
Thus, combining Eqs.\ (\ref{A_matrix}) and (\ref{|A|_matrix}), the unitary transformation of $U$ gives
\begin{equation}\label{U_matrix}
\check{U}\equiv\mathbf{N}^{\frac{1}{2}}U\mathbf{N}^{-\frac{1}{2}}=\check{|A|}^{-1}\check{A}
=\left[
\begin{array}{cc}
0 & |\Lambda|^{-\frac{1}{2}}\\
-|\Lambda|^{\frac{1}{2}} & 0
\end{array}
\right],
\end{equation}
since $\textrm{Dom}(|\Lambda|)\subset\textrm{Dom}(|\Lambda|^{\frac{1}{2}})$. Finally, to arrive at an expression for
$\tilde{\mu}$, we combine Eqs.\ (\ref{mu_definition}) and (\ref{|A|_matrix}) according to Eq.\
(\ref{new_bilinear_form}) and obtain
\begin{equation}\label{mu_tilde_unstable_field}
\tilde{\mu}(\psi_1,\psi_2)=\langle\mathbf{N}^{\frac{1}{2}}\psi_1,\tilde{\mathbf{M}}\mathbf{N}^{\frac{1}{2}}
\psi_2\rangle_ { L^2\oplus L^2},
\end{equation}
with
\begin{equation}\label{M_tilde_matrix}
\tilde{\mathbf{M}}\equiv\frac{1}{2}\left[
\begin{array}{cc}
|\Lambda|^{\frac{1}{2}} & 0\\
0 & |\Lambda|^{-\frac{1}{2}}
\end{array}
\right].
\end{equation}
Therefore, by defining the real Hilbert spaces $\tilde{\mathcal{A}}_1$ and $\tilde{\mathcal{A}}_2$ as the completion of
$C_0^\infty(\Sigma_0)$ in the norms given by $\frac{1}{2}\langle\;\cdot\;, |\Lambda|^{\frac{1}{2}}\;\cdot\;
\rangle_{L^2}$ and $\frac{1}{2}\langle\;\cdot\;, |\Lambda|^{-\frac{1}{2}}\;\cdot\;\rangle_{L^2}$, respectively, one has
that $\tilde{\mathcal{A}}$ is isomorphic to $\tilde{\mathcal{A}}_1\oplus \tilde{\mathcal{A}}_2$.
 
In $\mathcal{S}$ the complex structure $J$ coincides with $U$, so we can use Eqs.\ (\ref{J_spectral_form}) and
(\ref{U_matrix}), together with $\mathbb{I}=P_++P_-$, to obtain
\begin{equation}\label{K_matrix}
\check{K}=\frac{\mathbb{I}+i\check{U}}{2}=\frac{1}{2}\left[
\begin{array}{cc}
1 & i|\Lambda|^{-\frac{1}{2}}\\
-i|\Lambda|^{\frac{1}{2}} & 1
\end{array}
\right].
\end{equation}
Equations (\ref{mu_tilde_unstable_field}) and (\ref{K_matrix}) allow us to write the inner product of
$\mathcal{H}$, defined in Eq. (\ref{kg_inner_product}), as
\begin{eqnarray*}
\langle
K\psi_1,K\psi_2\rangle_\mathcal{H}=\phantom{1111111111111111111111111111111}\nonumber\\
\left\langle\frac{Xf_1+iX^{\dagger-1}p_1}{\sqrt{2}},
\frac{Xf_2+iX^{\dagger-1}p_2}{\sqrt{2}}\right\rangle_{L^2_\mathbb{C}},
\end{eqnarray*}
where $\langle\;\cdot\;,\;\cdot\;\rangle_{L^2_\mathbb{C}}$ stands for the inner product of $L^2_\mathbb{C}$, the
complexified version of $L^2$. The closed operator $X$ is defined in such a way that
\begin{equation*}\label{X_definition}
X^\dagger X=\|\varkappa\|^{-\frac{1}{2}}|\Lambda|^{\frac{1}{2}}\|\varkappa\|^{-\frac{1}{2}}.
\end{equation*}
Therefore, we are led to also identify $\mathcal{H}$ with $L^2_\mathbb{C}$ and implement the map $K$ through the linear
map $k:\mathcal{P}\rightarrow L^2_\mathbb{C}$ given by
\begin{equation*}\label{L_square_representation}
(f,p)\mapsto \frac{Xf+iX^{\dagger-1}p}{\sqrt{2}}.
\end{equation*}
There are several unitarily equivalent identifications between $\mathcal{H}$ and $L^2_\mathbb{C}$. For instance, we
could have chosen
\begin{equation*}
X=|\Lambda|^{\frac{1}{4}}\|\varkappa\|^{-\frac{1}{2}}, 
\end{equation*}
which coincides with the representation found by Kay in Ref.\ \cite{kay_1978} for stable fields in static spacetimes
\cite{footnote_3}.

\subsection{Time translation for ``one-particle states''}
\label{sec:time_translation_in_H}
%%%%%%%%%%%%%%%%%%%%%%%%%%%%%%%%%%%%%%%%%%%%%%%%%%

Our next task is to implement in $\mathcal{F}$ the time translation defined over $(\mathcal{M},g_{ab})$ by $\alpha_t$.
This is accomplished through the construction of a strongly continuous unitary group $\mathcal{U}(t)$ satisfying
\begin{equation}\label{U_operator}
\mathcal{U}(t)W(\psi)\mathcal{U}(-t)=W(\tau_t\psi).
\end{equation}
Thus, to build such operator we have to investigate first how the time translation acts on $\mathcal{H}$.

We start by analyzing the continuity of $\tau_t$ in $\mathcal{P}$, according to the norm of $\mathcal{A}$.
Using the fact that $|\Lambda|$ is a closed operator and that $\tau_t$ maps $\mathcal{P}$ into itself, we obtain
\begin{equation}\label{continuity_strong_topology}
\lim_{t\rightarrow0}\|(\tau_t-\mathbb{I})\psi\|^2_{\mathcal{A}}=0,
\end{equation}
since $\lim_{t\rightarrow0}(\tau_t\psi-\psi)=0$. So, $\tau_t$ is strongly continuous. Consequently,
this map is also strongly continuous according to the norm of $\tilde{\mathcal{A}}$, due to the boundedness of $|A|$.
Furthermore, we also have  
\begin{equation}\label{eq_strong_topology}
\lim_{\delta\rightarrow0}
\left\|\pp{\frac{\tau_{t+\delta}-\tau_t}{\delta}+\mathbf{h}_0\tau_t}\psi\right\|^2_{\mathcal{A}}=0,
\end{equation}
due to Eq.\ (\ref{field_dynamics_matrix_form}). Thus, that equation holds in the strong sense both in $\mathcal{A}$ and
$\tilde{\mathcal{A}}$.

As mentioned above, the behavior of the bilinear form (\ref{mu_definition}) under time translations is
quite different whether we restrict ourselves to the stable or the unstable sector. Thus, it is convenient to study each
case separately. Consider the spectral projection $Q_B$ of $\Lambda$, associated with the set $B\subset\mathbb{R}$. We
introduce on $\mathcal{P}$ the operators $Q_\pm$, which are implemented through
\begin{equation}\label{spectral_projections}
\check{Q}_\pm\equiv\mathbf{N}^{\frac{1}{2}}Q_\pm\mathbf{N}^{-\frac{1}{2}}=\left[
\begin{array}{cc}
Q_{I^\pm} & 0\\
0 & Q_{I^\pm}
\end{array}
\right],
\end{equation}
with $I^+\equiv[0,+\infty)$ and $I^-\equiv(-\infty,0)$. Since they are built out of spectral projections of $\Lambda$,
these operators define orthogonal projections on $\mathcal{A}$, $\tilde{\mathcal{A}}$, and $\mathcal{H}$. Hence, these
Hilbert spaces are isomorphic to the Hilbert spaces $\mathcal{A}^+\oplus\mathcal{A}^-$, $\tilde{\mathcal{A}}^+\oplus
\tilde{\mathcal{A}}^-$, and $\mathcal{H}^+\oplus \mathcal{H}^-$, respectively, where we have defined
$\mathcal{A}^\pm\equiv Q_\pm\mathcal{A}$, $\tilde{\mathcal{A}}^\pm\equiv Q_\pm\tilde{\mathcal{A}}$, and
$\mathcal{H}^\pm\equiv Q_\pm\mathcal{H}$ . 

Denote by $\mathcal{P}_\pm$ the images of $\mathcal{P}$ by the projections $Q_\pm$. In general, we do not expect
$\mathcal{P}_\pm$ to be subsets of $\mathcal{P}$. Thus, it is more convenient to define the time translation on
$\mathcal{P}_\pm$ as
\begin{equation}\label{time_translation_pm}
\tau_t^\pm Q_\pm\psi\equiv Q_\pm\tau_t\psi.
\end{equation}
Taking the derivative in both sides of Eq.\ (\ref{time_translation_pm}), using Eq.\ (\ref{field_dynamics_matrix_form}),
and the closure of $\mathbf{h}_0$ in $L^2\oplus L^2$, $\mathbf{h}$, we get
\begin{eqnarray*}
\frac{d}{dt}\tau_t^\pm Q_\pm\psi&=&Q_\pm\frac{d}{dt}\tau_t\psi\\
&\stackrel{\tr{s}}{=}&-Q_\pm\mathbf{h}_0\psi\\
&=&-Q_\pm\mathbf{h}\psi\\
&=&-\mathbf{h}Q_\pm\psi.
\end{eqnarray*}
Then, by defining $\mathbf{h}_\pm$ as the restriction of $\mathbf{h}$ to $\mathcal{P}_\pm$ it is possible to write
\begin{equation}\label{tau_pm_derivative}
\frac{d}{dt}\tau_t^\pm\stackrel{\tr{s}}{=}-\mathbf{h}_\pm\tau_t^\pm,
\end{equation}
where the equality holds in the strong sense both in $\mathcal{A}$ and $\tilde{\mathcal{A}}$.

In order to analyze the action of the time translation on the stable sector, we first note that
\begin{equation}\label{h_plus_skew_symmetric}
\langle\mathbf{h}_+\psi_1,\psi_2\rangle_{\mathcal{A}}=-\langle\psi_1,\mathbf{h}_+\psi_2\rangle_{\mathcal{A}},
\end{equation}
since $|\Lambda|$ and $\Lambda$ coincide in $\mathcal{P}_+$. Then, making use of Eq.\ (\ref{tau_pm_derivative}),
\begin{eqnarray}\label{energy_conservation}
\frac{d}{dt}\|\tau_t^+\psi\|^2_{\mathcal{A}}&=&-\langle\mathbf{h}_+\tau_t^+\psi,\tau_t^+\psi\rangle_{\mathcal{A}}
-\langle\tau_t^+\psi,\mathbf{h}_+\tau_t^+\psi\rangle_{\mathcal{A}}\nonumber\\
&=&0,
\end{eqnarray}
thanks to Eq.\ (\ref{h_plus_skew_symmetric}). This result is just a manifestation of the bilinear form $\mu$ coinciding
with the conserved energy in the stable sector. In conclusion, $\tau_t^+$ defines a isometry on $\mathcal{P}_+$
according the norm of $\mathcal{A}$. Furthermore, insofar as $\tau_t$ and $|A|$ commute, $\tau_t^+$ is also an
isometry according the norm of $\tilde{\mathcal{A}}$. Hence, the extension of $\tau_t^+$ to
$\tilde{\mathcal{A}}^+_\mathbb{C}$ defines a strongly continuous one-parameter unitary group which we denote by
$T_+(t)$. In the complex Hilbert space $\tilde{\mathcal{A}}^+_\mathbb{C}$, $\mathbf{h}_+$ is a skew-symmetric, densely
defined operator which maps $\mathcal{P}_++i\mathcal{P}_+$ into itself. Besides, $\mathbf{h}_+$ also commutes with
$T_+(t)$. So, following the strategy of Ref.\ \cite{kay_1978}, it is possible to use a lemma due to Nelson (see Lemma
10.1 of Ref.\ \cite{nelson_1959}; see also Lemma 2.1 of Ref.\ \cite{chernoff_1973}) to conclude that, as an operator on
$\tilde{\mathcal{A}}^+_\mathbb{C}$, $\mathbf{h}_+$ is essentially skew-adjoint in $\mathcal{P}_++i\mathcal{P}_+$. Then,
thanks to Stone's theorem and Eq.\ (\ref{tau_pm_derivative}),
\begin{equation}\label{T_plus_exp_form}
T_+(t)=e^{-th},
\end{equation}
where $h$ stands for the closure of $\mathbf{h}_+$ in $\tilde{\mathcal{A}}^+_\mathbb{C}$. The operator $T_+(t)$ can
be represented as
\begin{equation}\label{T_plus_matrix_form}
T_+(t)=\left[
\begin{array}{cc}
\cos(t|\Lambda|^{\frac{1}{2}}) & |\Lambda|^{-\frac{1}{2}}\sin(t|\Lambda|^{\frac{1}{2}})\\
-|\Lambda|^{\frac{1}{2}}\sin(t|\Lambda|^{\frac{1}{2}}) & \cos(t|\Lambda|^{\frac{1}{2}})
\end{array}
\right]
\end{equation}
and its restriction to $\mathcal{H}^+$ gives
\begin{equation}\label{T_plus_in_H}
T_+(t)\upharpoonright \mathcal{H}^+=\left[
\begin{array}{cc}
e^{-it|\Lambda|^{\frac{1}{2}}} & 0\\
0 & e^{-it|\Lambda|^{\frac{1}{2}}}
\end{array}
\right].
\end{equation}

As for the unstable sector, we encounter
\begin{equation}\label{h_minus_symmetric}
\langle\mathbf{h}_-\psi_1,\psi_2\rangle_{\mathcal{A}}=\langle\psi_1,\mathbf{h}_-\psi_2\rangle_{\mathcal{A}},
\end{equation}
since now $\Lambda$ coincides with $-|\Lambda|$ in $\mathcal{P}_-$. Then, from Eqs.\ (\ref{tau_pm_derivative}) and
(\ref{h_minus_symmetric}),
\begin{eqnarray}\label{bounded_tau_minus_second_derivative}
\frac{d^2}{dt^2}\|\tau_t^-\psi\|_{\mathcal{A}}^2&=&
4\|\mathbf{h}_-\tau_t^-\psi\|_{\mathcal{A}}^2\nonumber\\
&\leq&4M_2\epsilon_2\|\tau_t^-\psi\|_{\mathcal{A}}^2,
\end{eqnarray}
where we have used Eq.\ (\ref{bounded_killing_vector_norm}) and property (\ref{spectral_condition_2}) of
$\sigma(\Delta)$ to obtain the inequality above. Equation (\ref{bounded_tau_minus_second_derivative}) tell us that
there is a strictly positive function $C(t)$ such that
\begin{equation}\label{tau_minus_bounded}
\|\tau_t^-\psi\|_{\mathcal{A}}^2\leq C(t) \|\psi\|_{\mathcal{A}}^2.
\end{equation}
Equation (\ref{tau_minus_bounded}) establishes that $\tau_t^-$ is bounded in $\mathcal{A}^-$ and, consequently, also
in $\tilde{\mathcal{A}}^-$. Hence, the extension of $\tau_t^-$ to $\tilde{\mathcal{A}}^-_\mathbb{C}$ defines a
strongly continuous one-parameter bounded group which we denote by $T_-(t)$. In $\tilde{\mathcal{A}}^-_\mathbb{C}$,
$\mathbf{h}_-$ is a symmetric bounded operator. So, we can write $T_-(t)=e^{-t\mathbf{h}_-}$ and represent it as 
\begin{equation}\label{T_minus_matrix_form}
T_-(t)=\left[
\begin{array}{cc}
\cosh(t|\Lambda|^{\frac{1}{2}}) & |\Lambda|^{-\frac{1}{2}}\sinh(t|\Lambda|^{\frac{1}{2}})\\
|\Lambda|^{\frac{1}{2}}\sinh(t|\Lambda|^{\frac{1}{2}}) & \cosh(t|\Lambda|^{\frac{1}{2}})
\end{array}
\right].
\end{equation}
When restricted to $\mathcal{H}^-$, this representation leads to
\begin{eqnarray}\label{T_minus_in_H}
T_-(t)\upharpoonright \mathcal{H}^-=\phantom{111111111111111111111111111111111}\nonumber\\
\sqrt{2}\left[
\begin{array}{cc}
\cosh(t|\Lambda|^{\frac{1}{2}}-\frac{i\pi}{4}) & 0\\
0 & i\sinh(t|\Lambda|^{\frac{1}{2}}-\frac{i\pi}{4})
\end{array}
\right].\phantom{11}
\end{eqnarray}

In conclusion, we have managed to show that $\tau_t$ extends to $\tilde{\mathcal{A}}_\mathbb{C}=
\tilde{\mathcal{A}}_\mathbb{C}^+\oplus \tilde{\mathcal{A}}_\mathbb{C}^-$ as a strongly continuous one-parameter group
$T(t)$, which can be expressed as
\begin{equation}\label{T_operator}
T(t)=T_+(t)\oplus T_-(t).
\end{equation}
This family does not act as a unitary group on $\tilde{\mathcal{A}}_\mathbb{C}$ and does not commute with the
projections $P_\pm$, except when restricted to $\tilde{\mathcal{A}}_\mathbb{C}^+$. Hence, in general, $T(t)$ will map
the elements of $\mathcal{H}$ into $\mathcal{H}\oplus \overline{\mathcal{H}}$, ``mixing'' solutions that initially had
positive norm with ``negative-norm'' ones, according to the notion established on $\Sigma_0$.

% Before changing the subject, it is worth pointing out that part of the purpose of the representation set by Eq.\
% (\ref{mu_definition}) is the implementation of the time evolution for the stable and unstable sectors of $\mathcal{H}$
% as in Eqs.\ (\ref{T_plus_in_H}) and (\ref{T_minus_in_H}). In this sense, then, the representation of $\mathcal{H}$ in
% terms of $L^2_\mathbb{C}$ --- see discussion below Eq.\ (\ref{K_matrix}) --- is useless when the field is unstable,
% since Eq.\ (\ref{T_minus_in_H}) prevents us from representing $T(t)$ as an operator on this space.

\subsection{Time translation on the Fock space}
\label{sec:time_translation_in_F}
%%%%%%%%%%%%%%%%%%%%%%%%%%%%%%%%%%%%%%%%%%%%%%%%%%
 
Now, let us get back to the time translation operator acting on the space of states of the quantum field. Equation
(\ref{U_operator}) is equivalent to
\begin{equation}\label{time_translated_field}
\mathcal{U}(t)F_0(\psi)\mathcal{U}(-t)=ia[P_-T(t)\psi]-ia^*[P_+T(t)\psi]
\end{equation}
on $\textrm{span}(\mathcal{F}_0)$. The time translation ``mixes'' positive- and negative-norm solutions when the field
is unstable just like in an ``in-out'' setting the $S$-matrix maps the ``in'' space of states into the ``out'' one.
Thus, here we shall employ the approach presented in Ref.\ \cite{wald_1979} to define the action of $\mathcal{U}(t)$ on
$\mathcal{F}$.

The first step consists in setting the action of $\mathcal{U}(t)$ on $\Psi_0$. We start by defining the one-parameter
families of bounded operators $A_t:\mathcal{H}\rightarrow\mathcal{H}$ and
$B_t:\mathcal{H}\rightarrow\overline{\mathcal{H}}$ as 
\begin{equation}\label{A_t_operator}
A_t\equiv P_+T(t)\upharpoonright\mathcal{H} 
\end{equation}
and
\begin{equation}\label{B_t_operator}
B_t\equiv P_-T(t)\upharpoonright\mathcal{H}.
\end{equation}
From  Eq.\ (\ref{kg_inner_product}) and the conservation of the symplectic form (\ref{symplectic_form}), it follows that
for $\psi_1,\psi_2\in\mathcal{H}$
\begin{eqnarray*}
\langle\psi_1,\psi_2\rangle_\mathcal{H}&=&i\Omega^{\prime\prime}(\overline{\psi}_1,\psi_2)\\
&=&i\Omega^{\prime\prime}[\overline{T(t)\psi}_1,T(t)\psi_2]\\
&=&i\Omega^{\prime\prime}[\overline{(P_++P_-)T(t)\psi}_1,(P_++P_-)T(t)\psi_2]\\
&=&\langle A_t\psi_1,A_t\psi_2\rangle_\mathcal{H}-\langle B_t\psi_1,B_t\psi_2\rangle_{\overline{\mathcal{H}}}\\
&=&\langle \psi_1,({A_t}^\dagger A_t-{B_t}^\dagger B_t)\psi_2\rangle_\mathcal{H}
\end{eqnarray*}
and that
\begin{eqnarray*}
\langle\overline{\psi}_1,\psi_2\rangle_\mathcal{H}&=&i\Omega^{\prime\prime}(\psi_1,\psi_2)\\
&=&i\Omega^{\prime\prime}[(P_++P_-)T(t)\psi_1,(P_++P_-)T(t)\psi_2]\\
&=&\langle \overline{B_t\psi}_1,A_t\psi_2\rangle_\mathcal{H}-
\langle \overline{A_t\psi}_1,B_t\psi_2\rangle_{\overline{\mathcal{H}}}\\
&=&\langle({A_t}^\dagger\overline{B}_t-{B_t}^\dagger\overline{A}_t)
\overline{\psi}_1,\psi_2\rangle_\mathcal{H}\\
&=&0.
\end{eqnarray*}
Here, the operators $\overline{A}_t$ and $\overline{B}_t$ are defined as
\begin{equation*}
\overline{A}_t\equiv P_-T(t)\upharpoonright\overline{\mathcal{H}} 
\end{equation*}
and
\begin{equation*}
\overline{B}_t\equiv P_+T(t)\upharpoonright\overline{\mathcal{H}}.
\end{equation*}
Hence, the operators $A_t$ and $B_t$ satisfy
\begin{equation}\label{bogoliubov_transformation_1}
{A_t}^\dagger A_t-{B_t}^\dagger B_t=\mathbb{I}
\end{equation}
and
\begin{equation}\label{bogoliubov_transformation_2}
{A_t}^\dagger\overline{B}_t={B_t}^\dagger\overline{A}_t.
\end{equation}
Equations (\ref{bogoliubov_transformation_1}) and (\ref{bogoliubov_transformation_2}) show that the operators $A_t$ and
$B_t$ define a one-parameter family of Bogoliubov transformations. Furthermore, from Eq.\
(\ref{bogoliubov_transformation_1}) it is possible to show that $A_t$ is bounded from below. Hence ${A_t}^{-1}$ exists
as a bounded operator.

Next, write  
\begin{equation}\label{time_translated_vacuum}
\Psi_t\equiv\mathcal{U}(t)\Psi_0=c_t\left(1,\psi^{(1)}_t,\psi^{(2)}_t,\psi^{(3)}_t,\dots\right).
\end{equation}
By choosing $\psi=\overline{{A_t}^{-1}\chi}$ in Eq.\ (\ref{time_translated_field}) and applying its left-hand side on
$\mathcal{U}(t)\Psi_0$, we obtain
\begin{equation}\label{S_matrix_vacuum_eq}
0=a(\overline{\chi})\Psi_t-a^*(\varepsilon_t\overline{\chi})\Psi_t,
\end{equation}
with
\begin{equation}\label{two_particle_state}
\varepsilon_t\equiv\overline{B_t{A_t}^{-1}}.
\end{equation}
Equation (\ref{S_matrix_vacuum_eq}) fixes the action of $\mathcal{U}(t)$ on $\Psi_0$. This equation implies that
all $\psi^{(n)}_t$ in Eq.\ (\ref{time_translated_vacuum}) are null for $n$ odd, while for $n$ even they can be expressed
in terms of symmetrized tensor products of $\psi^{(2)}_t$. Hence, the only equation we must care about is
\begin{equation}\label{S_matrix_recurrence_relations_2}
\overline{\chi}\cdot\psi^{(2)}_t=\frac{1}{\sqrt{2}}\varepsilon_t\overline{\chi}.
\end{equation}
Following Ref.\ \cite{wald_1979}, we interpret $\psi_t^{(2)}$ in the left-hand side of Eq.\
(\ref{S_matrix_recurrence_relations_2}) as a map from $\overline{\mathcal{H}}$ to $\mathcal{H}$. Then, the
identification of $\psi^{(2)}_t$ with the operator $\varepsilon_t$ is consistent if the former is symmetric, i.e., if
\begin{equation}\label{epsilon_symmetry}
{\overline{\varepsilon}_t}^\dagger=\varepsilon_t,
\end{equation}
and if it is in the Hilbert-Schmidt class, meaning that
\begin{equation}\label{epsilon_hilbert_schmidt_class}
\textrm{tr}({\varepsilon_t}^\dagger\varepsilon_t)<+\infty.
\end{equation}
The first condition comes from the symmetry of $\psi_t^{(2)}$, while the second is a consequence of
$\|\psi_t^{(2)}\|_{\mathcal{H}\otimes\mathcal{H}}<+\infty$. Equation (\ref{epsilon_symmetry}) can be verified directly
from Eq.\ (\ref{bogoliubov_transformation_2}). As for Eq.\ (\ref{epsilon_hilbert_schmidt_class}), it gives a condition
for the existence of a solution of Eq.\ (\ref{S_matrix_vacuum_eq}). This condition is equivalent to
\begin{equation}\label{B_hilbert_schmidt_class}
\textrm{tr}({B_t}^\dagger B_t)<+\infty,
\end{equation}
since ${A_t}^{-1}$ is bounded. This condition can be read as stating that the expected value of the total number
operator in the state $\Psi_t$ is finite --- see Eq.\ (\ref{quanta_creation}) below. In conclusion, we have obtained
that
$\mathcal{U}(t)$ maps $\Psi_0$ into
\begin{equation}\label{time_translated_vacuum_state}
\Psi_t=c_t\left(1,0,\frac{1}{\sqrt{2}}\varepsilon_t,\dots,\frac{\sqrt{n!}}{2^{n/2}(\frac{n}{2})!}\bigotimes^{n/2}
\varepsilon_t, \dots\right),
\end{equation}
for $n$ even, provided that Eq.\ (\ref{B_hilbert_schmidt_class}) holds.

The definition of the time translation operator $\mathcal{U}(t)$ is completed by stating its action on $\mathcal{F}_0$,
the set of all $n$-particle states. Since $\textrm{span}(\mathcal{F}_0)$ is a dense subset of $\mathcal{F}$, this
defines $\mathcal{U}(t)$ on a dense domain. For a collection $\chi_1,\chi_2,\dots,\chi_n$ of vectors in $\mathcal{H}$,
the correspondent $n$-particle state is given by $\prod_{j=1}^n a^*(\chi_j)\Psi_0$. The action of
$\mathcal{U}(t)$ on these vectors is given by
\begin{equation}\label{time_translation_n_particle_states}
\mathcal{U}(t)\prod_{j=1}^n
a^*(\chi_j)\Psi_0=\prod_{j=1}^n\left[a^*(A_t\chi_j)-a(B_t\chi_j)\right]\Psi_t.
\end{equation}

It is left to check that the operator $\mathcal{U}(t)$ as constructed in the last paragraphs actually defines a
strongly continuous one-parameter unitary group on $\mathcal{F}$. The first thing to check is whether the norm of
$\Psi_t$ is finite. As shown in Ref.\ \cite{wald_1979}, assuming that $\varepsilon_t$ is Hilbert-Schmidt and using its
symmetry property, it is possible to prove that $\|\Psi_t\|_\mathcal{F}<+\infty$. Besides, the vector $\Psi_t$ lies in
the domain of powers of creation and annihilation operators. So, the right-hand side of Eq.\
(\ref{time_translation_n_particle_states}) is well-defined. To show that $\mathcal{U}(t)$ is unitary, it is enough to
verify that this operator preserves the inner product in $\mathcal{F}_0$. The inner product between two $n$-particle
states can be written in terms of the commutator (\ref{annihilation_creation_commutator}); since this commutator is
preserved by $\mathcal{U}(t)$, thanks to Eq.\ (\ref{bogoliubov_transformation_1}), the time translation defines an
unitary operator in a dense domain and, consequently, in $\mathcal{F}$. To see that $\mathcal{U}(t)$ inherits
the group property of $T(t)$, first we need the following relations, derived from Eqs.\ (\ref{A_t_operator}) and
(\ref{B_t_operator}):
\begin{equation}\label{relation_1}
A_{t_2}A_{t_1}+\overline{B_{t_2}\overline{B_{t_1}}}=A_{t_1+t_2}
\end{equation}
and
\begin{equation}\label{relation_2}
B_{t_2}A_{t_1}+\overline{A_{t_2}\overline{B_{t_1}}}=B_{t_1+t_2}.
\end{equation}
Equations (\ref{relation_1}) and (\ref{relation_2}), together with Eq.\ (\ref{S_matrix_vacuum_eq}), lead to
\begin{equation}\label{vacuum_group}
\mathcal{U}(t_2)\Psi_{t_1}=\Psi_{t_1+t_2}.
\end{equation}
Then, Eqs.\ (\ref{relation_1}) - (\ref{vacuum_group}) imply that
\begin{eqnarray*}
\mathcal{U}(t_2)\mathcal{U}(t_1)\prod_{j=1}^na^*(\chi_j)\Psi_0&=&
\prod_{j=1}^n\left[a^*(A_{t_1+t_2}\chi_j)\right.\\
&&\left.-a(B_{t_1+t_2}\chi_j)\right]\Psi_{t_1+t_2},
\end{eqnarray*}
verifying that $\mathcal{U}(t)$ is a one-parameter group. For the strong continuity of $\mathcal{U}(t)$, it is
sufficient to prove that
\begin{equation}\label{time_translation_strong_continuity}
\lim_{t\rightarrow0}\|\mathcal{U}(t)\prod_{j=1}^na^*(\chi_j)\Psi_0-
\prod_{j=1}^na^*(\chi_j)\Psi_0\|^2_\mathcal{F}=0.
\end{equation}
The form of the right-hand side of Eq.\ (\ref{time_translation_n_particle_states}) and the fact that the creation and
annihilation operators are continuous tell us that Eq.\ (\ref{time_translation_strong_continuity}) holds if (i) $A_t$
and $B_t$ are strongly continuous and if (ii) $\lim_{t\rightarrow0}\Psi_t=\Psi_0$. Condition (i) follows from the
definition of operators $A_t$ and $B_t$, Eqs.\ (\ref{A_t_operator}) and (\ref{B_t_operator}), and the fact that $T(t)$
is strongly continuous. Condition (ii) depends on the norm of $\varepsilon_t$ converging to zero as $t\rightarrow0$. But
\begin{equation*}
\lim_{t\rightarrow0}\|\varepsilon_t\|^2_{\mathcal{H}\otimes\mathcal{H}}\leq
\lim_{t\rightarrow0}\sum_{j=1}^\infty\|\overline{B_t\varphi_j}\|^2_\mathcal{H}=0,
\end{equation*}
where $\{\varphi_j\}_{j\in\mathbb{N}^*}$ is a complete orthonormal base of $\mathcal{H}$ and we have used Eqs.\
(\ref{B_t_operator}) and (\ref{B_hilbert_schmidt_class}).

It is possible to extract more information about the time-translated vacuum state $\Psi_t$ if we exploit the
representation of the operators $P_\pm$ and $T(t)$ when restricted to $\mathcal{H}^-$. Using Eqs.\ (\ref{K_matrix})
and (\ref{T_minus_matrix_form}), we get
\begin{eqnarray*}
P_\pm T_-(t)&=&\left[
\begin{array}{cc}
\cosh(t|\Lambda|^{\frac{1}{2}}) & 0\\
0 & \cosh(t|\Lambda|^{\frac{1}{2}})\\
\end{array}
\right]P_\pm\nonumber\\
&&\mp \left[
\begin{array}{cc}
i\sinh(t|\Lambda|^{\frac{1}{2}}) & 0\\
0 & i\sinh(t|\Lambda|^{\frac{1}{2}})
\end{array}
\right]R_\pm P_\mp,
\end{eqnarray*}
where
\begin{equation*}
R_\pm=\pm \left[
\begin{array}{cc}
0 & i|\Lambda|^{-\frac{1}{2}}\\
i|\Lambda|^{\frac{1}{2}} & 0
\end{array}
\right].
\end{equation*}
We can use the relation above to obtain the following representation for the operators $A_t$ and $B_t$ on
$K\mathcal{P}+iK\mathcal{P}\subset\mathcal{H}^+\oplus\mathcal{H}^-$:
\begin{equation}\label{A_t_matrix}
A_t={A_t}^+\oplus {A_t}^-,
\end{equation}
with
\begin{equation}\label{A_t_plus_matrix}
A^+_t\equiv\left[
\begin{array}{cc}
e^{-it|\Lambda|^\frac{1}{2}} & 0\\
0 & e^{-it|\Lambda|^\frac{1}{2}}
\end{array}
\right]
\end{equation}
and
\begin{equation}\label{A_t_minus_matrix}
A^-_t\equiv\left[
\begin{array}{cc}
\cosh(t|\Lambda|^{\frac{1}{2}}) & 0\\
0 & \cosh(t|\Lambda|^{\frac{1}{2}})
\end{array}
\right],
\end{equation}
and
\begin{equation}\label{B_t_matrix}
B_t=B^+_t\oplus B^-_t,
\end{equation}
with 
\begin{equation}\label{B_t_plus_matrix}
B^+_t\equiv0 
\end{equation}
and
\begin{equation}\label{B_t_minus_matrix}
B^-_t\equiv \left[
\begin{array}{cc}
i\sinh(t|\Lambda|^{\frac{1}{2}}) & 0\\
0 & i\sinh(t|\Lambda|^{\frac{1}{2}})
\end{array}
\right]R_-.
\end{equation}
Combining Eqs.\ (\ref{A_t_matrix}) - (\ref{B_t_minus_matrix}) according to Eq.\ (\ref{two_particle_state}), we obtain
\begin{equation}\label{two_particle_state_matrix}
\varepsilon_t=\varepsilon^+_t\oplus \varepsilon^-_t,
\end{equation}
where 
\begin{equation}\label{two_particle_state_matrix_plus}
\varepsilon^+_t\equiv0
\end{equation}
and
\begin{equation}\label{two_particle_state_matrix_minus}
\varepsilon^-_t\equiv \left[
\begin{array}{cc}
i\tanh(t|\Lambda|^{\frac{1}{2}}) & 0\\
0 & i\tanh(t|\Lambda|^{\frac{1}{2}})
\end{array}
\right]R_+.
\end{equation}

Now we possess the necessary tools to calculate the mean value of the number operator in the state $\Psi_t$. Consider
again an orthonormal base $\{\varphi_j\}_{j\in\mathbb{N}^*}$ of $\mathcal{H}$ and define the operator $N(\varphi_j)$ as
\begin{equation*}\label{number_operator}
N(\varphi_j)\equiv a^*(\varphi_j) a(\overline{\varphi}_j),
\end{equation*}
the number operator associated to $\varphi_j$. The mean value of $N(\varphi_j)$ in the state $\Psi_t$ is given by
\begin{equation}\label{quanta_creation}
\langle
\Psi_t,N(\varphi_j)\Psi_t\rangle_\mathcal{F}=\langle\overline{B_t\varphi_j},
\overline{B_t\varphi_j}\rangle_\mathcal{H}.
\end{equation}
Note that, as stated earlier in this section, the expected value of the total number operator is finite if Eq.\
(\ref{B_hilbert_schmidt_class}) holds. Substituting Eqs.\ (\ref{B_t_matrix}) - (\ref{B_t_minus_matrix}) into Eq.\
(\ref{quanta_creation}) it follows that
\begin{eqnarray*}
\langle\Psi_t,N(\varphi_j)\Psi_t\rangle_\mathcal{F}&=&
\int_{-\infty}^0{|\lambda|^\frac{3}{2}\sinh^2(|\lambda|^\frac{1}{2}t)d\mu_j}\\
&&+\int_{-\infty}^0{|\lambda|^{-\frac{3}{2}}\sinh^2(|\lambda|^\frac{1}{2}t)d\nu_j},
\end{eqnarray*}
where
\begin{equation*}
d\mu_j\equiv d\langle\|\varkappa\|^{-\frac{1}{2}}f_j,Q_\lambda\|\varkappa\|^{-\frac{1}{2}}f_j\rangle_{L^2_\mathbb{C}}
\end{equation*}
and
\begin{equation*}
d\nu_j\equiv d\langle\|\varkappa\|^{\frac{1}{2}}p_j,Q_\lambda\|\varkappa\|^{\frac{1}{2}}p_j\rangle_{L^2_\mathbb{C}},
\end{equation*}
with $\varphi_j=(f_j,p_j)$, are spectral measures associated with the operator $\Lambda$. Thus, the mean number of
quanta created in a given ``one-particle state'' grows exponentially in time.

\subsection{The Hamiltonian operator}
\label{sec:hamiltonian}
%%%%%%%%%%%%%%%%%%%%%%%%%%%%%%%%%%%%%%%%%%%%%%%%%%  

In order to investigate further the time translation operator defined in the last section, it is convenient to separate
the space of states of the quantum field in two pieces, one describing only the stable degrees of freedom of the field
and another describing only the unstable ones. So, consider the Hilbert spaces $\mathcal{F}^\pm\equiv
\mathfrak{F}_\textrm{s}(\mathcal{H}^\pm)$. The spaces $\mathcal{F}$ and $\mathcal{F}^+\otimes\mathcal{F}^-$ are
isomorphic, according to the following map --- see, e.g., Ref.\ \cite{wald_1975}. Take $\eta^{(m)} \in
(\bigotimes^m\mathcal{H}^+)_\textrm{s}$, $\varrho^{(n)}\in(\bigotimes^n\mathcal{H}^-)_\textrm{s}$, and for each
$(m+n)$-particle state in $\mathcal{F}$ of the form
\begin{equation}\label{n_particle_state_F}
\pp{\dots,\eta^{(m)}\otimes_\textrm{s}\varrho^{(n)},\dots}
\end{equation}
associate the vector
\begin{equation}\label{n_particle_state_F_plus_times_F_minus}
\sqrt{\frac{m!n!}{(m+n)!}}\pp{\dots,\eta^{(m)},\dots}\otimes\pp{\dots,\varrho^{(n)},\dots}
\end{equation}
in $\mathcal{F}^+\otimes\mathcal{F}^-$. The combinatorial factor in Eq.\
(\ref{n_particle_state_F_plus_times_F_minus}), together with the fact that the contraction between $\eta^{(m)}$ and
$\varrho^{(n)}$ is null, makes this map norm preserving on these vectors. Furthermore, the set formed by all vectors of
the form (\ref{n_particle_state_F}) is dense in $\mathcal{F}$, whereas the set formed by all vectors of the form
(\ref{n_particle_state_F_plus_times_F_minus}) is dense in $\mathcal{F}^+\otimes\mathcal{F}^-$. Therefore, this map
establishes the desired isomorphism between these two Hilbert spaces. Denote by $\Psi_0^\pm$ the vacuum states of
$\mathcal{F}^\pm$. Applying this map to the time-translated vacuum state, Eq.\ (\ref{time_translated_vacuum_state}), we
see
that
\begin{equation}\label{time_translated_vacuum_decomposition}
\Psi_t\mapsto\Psi_0^+\otimes\Psi_t^-,
\end{equation}
where
\begin{equation}\label{tachyonic_time_translated_vacuum}
\Psi_t^-=c_t\left(1,0,\frac{1}{\sqrt{2}}\varepsilon_t^-,\dots,\frac{\sqrt{n!}}{2^{n/2}(\frac{n}{2})!}\bigotimes^{n/2}
\varepsilon_t^-, \dots\right),
\end{equation}
due to Eqs. (\ref{two_particle_state_matrix}) - (\ref{two_particle_state_matrix_minus}).

Next, for $\eta\in\mathcal{H}^+$ and $\varrho\in\mathcal{H}^-$, denote by $b(\overline{\eta})$ and $b^*(\eta)$ the
annihilation and creation operators on $\mathcal{F}^+$ and by $c(\overline{\varrho})$ and $c^*(\varrho)$ the 
same operators but on $\mathcal{F}^-$ --- defined in the same manner as the operators in Eqs.\
(\ref{annihilation_operator}) and (\ref{creation_operator}). As their counterparts on $\mathcal{F}$, the annihilation
and creation operators on $\mathcal{F}^\pm$ are defined on the linear span of $\mathcal{F}^\pm_0$, the set of all
$n$-particle states of $\mathcal{F}^\pm$. It is easy to check, then, that the action of the operator
$a(\overline{\chi})$ on vectors of the form (\ref{n_particle_state_F}) can  be written as the action of the operator
\begin{equation}\label{annihilation_operator_decomposition}
b(\overline{\eta})\otimes\mathbb{I}+\mathbb{I}\otimes c(\overline{\varrho})
\end{equation}
on vectors of the form (\ref{n_particle_state_F_plus_times_F_minus}), while the operator $a^*(\chi)$ acts as
\begin{equation}\label{creation_operator_decomposition}
b^*(\eta)\otimes\mathbb{I}+\mathbb{I}\otimes c^*(\varrho),
\end{equation}
for $\chi=\eta+\varrho$. Taking $\psi\in\mathcal{P}$, Eqs.\ (\ref{annihilation_operator_decomposition}) and
(\ref{creation_operator_decomposition}) imply for the field operator, Eq.\ (\ref{field_operator}), that
\begin{equation*}
F_0(\psi)\mapsto F^+_0(\psi^+)\otimes\mathbb{I}+\mathbb{I}\otimes F^-_0(\psi^-),
\end{equation*}
where $\psi=\psi^++\psi^-$, $\psi^\pm\in\mathcal{P}_\pm$,
\begin{equation}\label{stable_field_operator}
F^+_0(\psi^+)\equiv ib(P_-\psi^+)-ib^*(P_+\psi^+),
\end{equation}
and
\begin{equation}\label{unstable_field_operator}
F^-_0(\psi^-)\equiv ic(P_-\psi^-)-ic^*(P_+\psi^-).
\end{equation}

Now, let us turn our attention to Eq.\ (\ref{time_translation_n_particle_states}). The action of creation and
annihilation operators on the states $\Psi_0$ and $\Psi_t$ results in vectors of the form (\ref{n_particle_state_F}).
Thus, if we write $A_t\chi_j=A^+_t\eta_j+A^-_t\varrho_j$, $B_t\chi_j=B^-_t\varrho_j$, and use the expressions
(\ref{annihilation_operator_decomposition}) and (\ref{creation_operator_decomposition}), we conclude from Eq.\
(\ref{time_translation_n_particle_states}) that the action of $\mathcal{U}(t)$ on $\mathcal{F}$ is equivalent to the
action of $\mathcal{V}(t)\otimes \mathcal{W}(t)$ on $\mathcal{F}^+\otimes \mathcal{F}^-$, where $\mathcal{V}(t)$ is
defined by
\begin{equation}\label{time_translation_n_particle_states_F_plus}
\mathcal{V}(t)\prod_{j=1}^nb^*(\eta_j)\Psi_0^+=\prod_{j=1}^nb^*(A^+_t\eta_j)\Psi_0^+,
\end{equation}
while $\mathcal{W}(t)$ is defined by
\begin{equation}\label{time_translation_n_particle_states_F_minus}
\mathcal{W}(t)\prod_{j=1}^nc^*(\varrho_j)\Psi_0^-=
\prod_{j=1}^n[c^*(A^-_t\varrho_j)-c(B^-_t\varrho_j)]\Psi_t^-.
\end{equation}
By the same arguments of the last section it is possible to show that, as $\mathcal{U}(t)$, the operators
$\mathcal{V}(t)$ and $\mathcal{W}(t)$ define strongly continuous one-parameter unitary groups on $\mathcal{F}^+$ and
$\mathcal{F}^-$, respectively. Note that Eq.\ (\ref{time_translation_n_particle_states_F_plus}) is equivalent to the
``second quantization'' of the restriction of $T_+(t)$ to $\mathcal{H}^+$, given in Eq.\ (\ref{T_plus_in_H}). 

The main goal of this section is the analysis of the generator of $\mathcal{W}(t)$. In particular, we wish to express it
in terms of the creation and annihilation operators defined in $\mathcal{F}^-$. For simplicity, we will make the
additional assumption that the instability is related exclusively to the point spectrum of $\Delta$ and that this
operator has finitely many eigenvalues. So, $\mathcal{H}^-$ is a finite-dimensional Hilbert space. This assumption
corresponds to the most common scenario explored in the literature where the scalar field is made unstable due to
a deep enough scalar potential well that allows bound states with ``imaginary frequencies.''

Thus, making use of Stone's theorem, write
\begin{equation}\label{time_translation_in_F_minus}
\mathcal{W}(t)=e^{-itH_-},
\end{equation}
where $H_-$ is a self-adjoint operator which maps $\textrm{span}(\mathcal{F}^-_0)$ into itself, thanks to Eq.\
(\ref{time_translation_n_particle_states_F_minus}). Applying the time translation to the field operator defined in Eq.\
(\ref{unstable_field_operator}) and taking the derivative with respect to $t$ at $t=0$, it follows that
\begin{equation}\label{Heisenberg_eq}
[H_-,F_0^-(\tau)]=c^*(P_+\mathbf{h}_-\tau)-c(P_-\mathbf{h}_-\tau).
\end{equation}
It is convenient to rewrite $P_\pm\mathbf{h}_-$ as
\begin{equation}\label{projector_time_derivative}
P_\pm\mathbf{h}_-=\mathbf{w}_\pm\mathbf{E}P_\mp,
\end{equation}
with
\begin{equation}\label{E_matrix}
\mathbf{E}\equiv\left[
\begin{array}{cc}
-1 & 0\\
0 & 1
\end{array}
\right].
\end{equation}
Next, taking advantage of the representation of $P_\pm$, Eq.\ (\ref{K_matrix}), and the form of $\mathbf{h}_-$ with the
transformation (\ref{unitary_transformation}),
\begin{equation*}
\check{\mathbf{h}}_-\equiv \mathbf{N}^{\frac{1}{2}}\mathbf{h}_-\mathbf{N}^{-\frac{1}{2}}=\left[
\begin{array}{cc}
0 & -1\\
\Lambda & 0
\end{array}
\right]=
-\left[
\begin{array}{cc}
0 & 1\\
|\Lambda| & 0
\end{array}
\right],
\end{equation*} 
we obtain
\begin{equation}\label{w_matrix}
\check{\mathbf{w}}_\pm\equiv\mathbf{N}^{\frac{1}{2}}\mathbf{w}_\pm\mathbf{N}^{-\frac{1}{2}}=\pm i\left[
\begin{array}{cc}
|\Lambda|^{\frac{1}{2}} & 0\\
0 & |\Lambda|^{\frac{1}{2}}
\end{array}
\right].
\end{equation}
From Eqs.\ (\ref{K_matrix}) and (\ref{E_matrix}) we can show that the combination $\mathbf{E}P_\pm$ satisfies
\begin{equation*}
P_\pm\mathbf{E}P_\pm=0.
\end{equation*}
Hence, $\mathbf{E}$ maps $\overline{\mathcal{H}}\phantom{}^-$ into $\mathcal{H}^-$ and vice-versa.
Denote by $\{\varphi_j\}_{j=1,\dots,N}$ the orthonormal base of $\mathcal{H}^-$ built out of the eigenvectors of
$\Delta$. Then, we can express $\tau$ as
\begin{equation}\label{unstable_solution}
\tau=\sum_{j=1}^N\alpha_j\varphi_j+\overline{\alpha}_j\overline{\varphi}_j.
\end{equation}
Substituting Eq.\ (\ref{unstable_solution}) in both sides of Eq.\ (\ref{Heisenberg_eq}) and using Eq.\
(\ref{unstable_field_operator}), we conclude that
\begin{eqnarray}\label{unstable_annihilation_operator_eq_motion}
[H_-,c(\overline{\varphi}_j)]&=&-ic^*(\mathbf{w}_+\mathbf{E}\overline{\varphi}_j)\nonumber\\
&=&|\lambda_j|^{\frac{1}{2}}c^*(\varphi_j)
\end{eqnarray}
and
\begin{eqnarray}\label{unstable_creation_operator_eq_motion}
[H_-,c^*(\varphi_j)]&=&-ic(\mathbf{w}_-\mathbf{E}\varphi_j)\nonumber\\
&=&-|\lambda_j|^{\frac{1}{2}}c(\overline{\varphi}_j),
\end{eqnarray}
where $\lambda_j$, for $j=1,\dots,N$, denote the negative eigenvalues of $\Delta$. Equations
(\ref{unstable_annihilation_operator_eq_motion}) and (\ref{unstable_creation_operator_eq_motion}) are
simultaneously satisfied if
\begin{equation}\label{unstable_hamiltonian}
H_-=-\sum_{j=1}^N\frac{|\lambda_j|^{\frac{1}{2}}}{2}[c(\overline{\varphi}_j)c(\overline{\varphi}_j)+
c^*(\varphi_j)c^*(\varphi_j)],
\end{equation}
on $\textrm{span}(\mathcal{F}^-_0)$, up to the addition of a multiple of the identity operator. Defining the operators
\begin{equation*}
\hat{P}_j\equiv -i\sqrt{\frac{|\lambda_j|^{\frac{1}{2}}}{2}}[c(\overline{\varphi}_j)-c^*(\varphi_j)]
\end{equation*}
and
\begin{equation*}
\hat{Q}_j\equiv \frac{1}{\sqrt{2|\lambda_j|^{\frac{1}{2}}}}[c(\overline{\varphi}_ j)+c^*(\varphi_j)],
\end{equation*}
it is possible to cast Eq.\ (\ref{unstable_hamiltonian}) as
\begin{equation}\label{upside_down_harmonic_oscillator}
H_-=\sum_{j=1}^N\frac{1}{2}\hat{P}_j^2-\frac{|\lambda_j|}{2}\hat{Q}_j^2.
\end{equation}
As expected, Eq.\ (\ref{upside_down_harmonic_oscillator}) tells us that each ``one-particle state'' $\varphi_j$ behaves
as a non-relativistic particle in a parabolic potential barrier.

%%%%%%%%%%%%%%%%%%%%%%%%%%%%%%%%%%%%%%%%%%%%%%%%%%%%%%%%%%%%%%%%%%%%%
\section{Conclusion and final remarks}
\label{sec:conclusion}
%%%%%%%%%%%%%%%%%%%%%%%%%%%%%%%%%%%%%%%%%%%%%%%%%%%%%%%%%%%%%%%%%%%%%

Throughout the text we have referred to $\mathcal{H}$ as a ``one-particle Hilbert space.'' In QFTCS there is no
preferred notion of particles by the theory, even though a natural one can be built in stationary spacetimes whether the
field is stable \cite{fulling_book,wald_red_book}. Despite of the fact that we have considered a static spacetime
background, the lack of a vacuum state invariant under time translations in the representation we have built leads to
the conclusion that one should not attribute a particle content to the system. As a matter of fact, it was shown that
particle detectors coupled to the field following the orbits of the time isometry copiously excite if the instability
analyzed here is present \cite{landulfo_lima_matsas_vanzella_2012}. This result is expected, since the Hamiltonian
operator of the quantum system is unbounded from below --- see Eq.\ (\ref{unstable_hamiltonian}) above.

As mentioned in the introductory part and implied by Eqs.\ (\ref{bounded_killing_vector_norm}) and
(\ref{spectral_condition_1}), here we have focused on unstable free scalar fields on static backgrounds without
horizons and possessing a ``mass gap.'' These assumptions are an essential part of the proof of the inequality
(\ref{bounded_symplectic_structure_mu}) for our choice for the bilinear form $\mu$, Eq.\ (\ref{mu_definition}) --- see
Appendix \ref{appendix}. Even tough restrictive, these conditions can be satisfied at least for massive scalar particles
in Minkowski spacetime subject to scalar potential wells \cite{schroer_and_swieca_1970} and for non-minimally coupled
scalar fields in static spacetimes curved by classical matter \cite{lima_and_vanzella_2010}. The former example
resembles the SSW instability, while the latter has a realization in the spacetime of compact objects
\cite{lima_matsas_vanzella_2010}.

In summary, here we have studied the quantization of a neutral scalar field that is made unstable through the coupling
with an external scalar potential while evolving on a static globally hyperbolic spacetime. We have shown how the
general formalism to find a one-particle Hilbert space based on the definition of a complex structure for linear
fields can be particularized in order to treat the instability. The key point of the mathematical construction is our
choice of the bilinear form given in Eq.\ (\ref{mu_definition}) which sets a representation of the Weyl relations for
the quantized field. Then, we have investigated the action of the time translation operator on the vacuum of this
representation and from this it was shown how to define its action on the whole space of states of the unstable quantum
field. As one would expect, the presence of the instability leads to the break of the invariance of the vacuum under
time translations. Finally, from the particular form that the dynamics assumes when specialized to the unstable sector
of the field, we have arrived at an expression for the contribution from the unstable degrees of freedom of the field to
the generator of the time translations, the Hamiltonian. The expression we have encountered tells us that each degree of
freedom lying in the unstable sector behaves just like a non-relativistic quantum particle subjected to a parabolic
potential barrier.

%%%%%%%%%%%%%%%%%%%%%%%%%%%%%%%%%%%%%%%%%%%%%%%%%%%%%%%%%%%%%%%%%%%%%
\acknowledgments

It is a pleasure to thank Daniel Vanzella for countless stimulating discussions concerning QFTCS and for his careful
reading of this manuscript. The author acknowledges financial support from the S\~ao Paulo Research Foundation (FAPESP)
under Grant No.\ 2012/00737-0.
%%%%%%%%%%%%%%%%%%%%%%%%%%%%%%%%%%%%%%%%%%%%%%%%%%%%%%%%%%%%%%%%%%%%%

\appendix

%%%%%%%%%%%%%%%%%%%%%%%%%%%%%%%%%%%%%%%%%%%%%%%%%%%%%%%%%%%%%%%%%%%%%
\section{}
\label{appendix}
%%%%%%%%%%%%%%%%%%%%%%%%%%%%%%%%%%%%%%%%%%%%%%%%%%%%%%%%%%%%%%%%%%%%%

In the present appendix we shall present the intermediate steps that lead to inequality
(\ref{bounded_symplectic_structure_mu}), for $\mu$ with the form assumed in Eq.\ (\ref{mu_definition}).

To prove the inequality (\ref{bounded_symplectic_structure_mu}) we first show that the symplectic form
(\ref{symplectic_form}) is bounded in the $L^2\oplus L^2$-norm when the spacetime is static \cite{footnote_2}. Thus,
for any $\psi_1,\psi_2\in\mathcal{P}$, 
\begin{eqnarray}\label{omega_bounded_L_square}
|\Omega(\psi_1,\psi_2)|^2
&=&\left|\int_{\Sigma_0}{d\Sigma\pp{p_1f_2-p_2f_1}}\right|^2\nonumber\\
&\le&\cc{\int_{\Sigma_0}{d\Sigma\pp{|p_1||f_2|+|p_2||f_1|}}}^2\nonumber\\
&\le&\left[\pp{\int_{\Sigma_0}{d\Sigma|p_1|^2}\int_{\Sigma_0}{d\Sigma|f_2|^2}}^{\frac{1}{2}}
\right.\nonumber\\
&\phantom{\le}&\left.+\pp{\int_{\Sigma_0}{d\Sigma|p_2|^2}
\int_{\Sigma_0}{d\Sigma|f_1|^2}}^{\frac{1}{2}}\right]^2\nonumber\\
&\le&\langle\psi_1,\psi_1\rangle_{L^2\oplus L^2}\langle\psi_2,\psi_2\rangle_{L^2\oplus L^2}.
\end{eqnarray}
For the second inequality in Eq.\ (\ref{omega_bounded_L_square}) we have applied the Cauchy-Schwartz inequality to the
inner product of $L^2$, while for the third we have used that
\begin{equation*}
|a||b|+|c||d|\le\sqrt{|a|^2+|d|^2}\sqrt{|b|^2+|c|^2}.
\end{equation*}
On the other hand, from property (\ref{spectral_condition_1}) of $\sigma(\Delta)$ and Eq.\
(\ref{bounded_killing_vector_norm}),
\begin{eqnarray}\label{mu_bounded_below_L_square}
\mu(\psi,\psi)
&\ge&\frac{1}{2}\int_{\Sigma_0}{d\Sigma\pp{\|\varkappa\||p|^2+M_1|f|^2}}\nonumber\\
&\ge&\frac{1}{2}\epsilon_1\textrm{min}(1,M_1/\epsilon_2)\langle\psi,\psi\rangle_{L^2\oplus L^2}.
\end{eqnarray}
Hence, combining Eqs.\ (\ref{omega_bounded_L_square}) and (\ref{mu_bounded_below_L_square}) we obtain, after rescaling
our temporal coordinate, that Eq.\ (\ref{bounded_symplectic_structure_mu}) holds when $\mu$ is given by Eq.\
(\ref{mu_definition}).


\begin{thebibliography}{}

\bibitem{lima_and_vanzella_2010} W.\ C.\ C.\ Lima and D.\ A.\ T.\ Vanzella, Phys.\ Rev.\ Lett.\ \textbf{104}, 161102
(2010).

\bibitem{lima_matsas_vanzella_2010} W.\ C.\ C.\ Lima, G.\ E.\ A.\ Matsas, and D.\ A.\ T.\ Vanzella, Phys.\ Rev.\ Lett.\
\textbf{105}, 151102 (2010).

\bibitem{lima_mendes_matsas_vanzella_2013} W.\ C.\ C.\ Lima, R.\ F.\ P.\ Mendes, G.\ E.\ A.\ Matsas, and D.\ A.\ T.\
Vanzella, Phys.\ Rev.\ D \textbf{87}, 104039 (2013).

\bibitem{pani_et_al_2011} P.\ Pani, V.\ Cardoso, E.\ Berti, J.\ Read, and M.\ Salgado, Phys. Rev. D \textbf{83}, 081501
(2011).

\bibitem{landulfo_lima_matsas_vanzella_2012} A.\ G.\ S.\ Landulfo, W.\ C.\ C.\ Lima, G.\ E.\ A.\ Matsas, and D.\ A.\ T.\
Vanzella, Phys.\ Rev.\ D \textbf{86}, 104025 (2012).

\bibitem{schiff_snyder_weinberg_1940} L.\ I.\ Schiff, H.\ Snyder, and J.\ Weinberg, Phys.\ Rev.\ \textbf{57}, 315
(1940).

\bibitem{fulling_1976} S.\ A.\ Fulling, Phys.\ Rev.\ D \textbf{14}, 1939 (1976).

\bibitem{schroer_and_swieca_1970} B.\ Schr\"oer and J.\ A.\ Swieca, Phys.\ Rev.\ D \textbf{2}, 2938 (1970).

\bibitem{fulling_book} S.\ A.\ Fulling, \textit{Aspects of Quantum Field Theory in Curved Space-Time} (Cambridge
University Press, Cambridge, England, 1989).

\bibitem{wald_red_book} R.\ M.\ Wald, \textit{Quantum Field Theory in Curved Spacetime and Black Hole Thermodynamics}
(University of Chicago Press, Chicago, 1994).

\bibitem{ashtekar_and_magnon_1975} A.\ Ashtekar and A.\ Magnon, Proc.\ R.\ Soc.\ A \textbf{346}, 375 (1975).

\bibitem{kay_1978} B.\ S.\ Kay, Comm.\ Math.\ Phys.\ \textbf{62}, 55 (1978).

\bibitem{chmielowski_1994} P.\ Chmielowski, Classical Quantum Gravity \textbf{11}, 41 (1994).

\bibitem{wald_1979} R.\ M.\ Wald, Ann.\ Phys.\ (N.Y.) \textbf{118}, 490 (1979).

\bibitem{bernal_and_sanchez_2003} A.\ N.\ Bernal and M.\ S\'anchez, Comm.\ Math.\ Phys.\ \textbf{243}, 461 (2003).

\bibitem{friedlander_book} F.\ G.\ Friedlander, \textit{The Wave Equation on a Curved Space-Time} (Cambridge University
Press, Cambridge, England, 1975).

\bibitem{leray_notes} J.\ Leray, \textit{Hyperbolic Partial Differential Equations} (Institute for Advanced Study,
Princeton, NJ, 1953).

\bibitem{segal_1962} I.\ E.\ Segal, Illinois J.\ Math.\ \textbf{6}, 500 (1962).

\bibitem{weinless_1969} M.\ Weinless, J.\ Func.\ Anal.\ \textbf{4}, 350 (1969).

\bibitem{reed_and_simon_v2} M.\ Reed and B.\ Simon, \textit{Methods of Modern Mathematical Physics}, vol.\ II (Academic,
New York, 1975).

\bibitem{haag_book} R.\ Haag, \textit{Local Quantum Physics: Fields, Particles, and Algebras} (Springer, Berlin, 1996).

\bibitem{reed_and_simon_v1} M.\ Reed and B.\ Simon, \textit{Methods of Modern Mathematical Physics}, vol.\ I (Academic,
New York, 1972).

\bibitem{kay_and_wald_1991} B.\ S.\ Kay and R.\ M.\ Wald, Phys.\ Rep.\ \textbf{207}, 49 (1991).

\bibitem{footnote_1} The extension of the symplectic form to $\tilde{\mathcal{A}}$, together with Eq.\
(\ref{lub_mu_tilde}), gives rise to a bounded operator $J^\prime$, which also endows $\tilde{\mathcal{A}}$ with a
complex structure. It turns out that $J=J^\prime$, since $U$ and $\Omega$ are both bounded and densely defined in
$\tilde{\mathcal{A}}$.

\bibitem{ikebe_and_kato_1962} T.\ Ikebe and T.\ Kato, Arch.\ Rational Mech.\ Anal.\ \textbf{9}, 77 (1962).

\bibitem{fulling_and_ruijsenaars_1987}  S.\ A.\ Fulling and S.\ M.\ N.\ Ruijsenaars, Phys.\ Rep.\ \textbf{152}, 135
(1987). 

\bibitem{wccl_phd_thesis} W.\ C.\ C.\ Lima, Ph.D.\ thesis, Instituto de F\'\i sica de S\~ao Carlos, Universidade de
S\~ao Paulo, 2012.

\bibitem{footnote_3} Unlike what was stated in the published version, the representation of the ``one-particle Hilbert space'' 
$\mathcal{H}$ in terms of $L^2_\mathbb{C}$ as discussed here is useful to represent the one-particle dynamics when the field is 
unstable. For the details, see the Erratum~\cite{erratum}. The rectification of that statement has no consequences for any of the 
other results presented in the paper.

\bibitem{nelson_1959} E.\ Nelson, Ann.\ Math.\ \textbf{70}, 572 (1959).

\bibitem{chernoff_1973} P.\ R.\ Chernoff, J.\ Func.\ Anal.\ \textbf{12}, 401 (1973).

\bibitem{wald_1975} R.\ M.\ Wald, Comm.\ Math.\ Phys.\ \textbf{45}, 9 (1975).

\bibitem{footnote_2} For our intentions in this study, proving this assertion is enough. However, it is worth pointing
out that even in non-stationary spacetimes the symplectic form (\ref{symplectic_form}) is bounded in the norm of
square-integrable functions in the measure $d\Sigma$. Hence, even in the lack of any other prescription, it is always
possible to construct a representation for the field algebra according to the recipe presented in Sec.\
\ref{sec:standard_quantization_general_spacetimes}.

\bibitem{erratum} W.~C.~C.~Lima, Phys.~Rev.~D \textbf{94}, 129901(E) (2016).

\end{thebibliography}
\end{document}